\documentclass[11pt]{article}
\usepackage{latexsym}
\usepackage{amsmath,amsbsy,amssymb}
\usepackage{verbatim}

\usepackage{rotating}
\usepackage{curves}
\usepackage{epic}
\usepackage{eepic}
\usepackage{epsfig}

\let\OLDthebibliography\thebibliography
\renewcommand\thebibliography[1]{
  \OLDthebibliography{#1}
  \setlength{\parskip}{0pt}
  \setlength{\itemsep}{0pt plus 0.3ex}
}

\newcommand{\dd}{\mbox{\rm d}}

\newcommand{\gam}{\gamma}
\newcommand{\Gam}{\Gamma}

\newcommand{\tl}{\tilde}

\newcommand{\ddt}{\frac{\dd}{\dd \tau}}
\newcommand{\hmu}{{\hat \mu}}
\newcommand{\dotx}{{\dot x}}
\newcommand{\lb}{\lbrace}
\newcommand{\rb}{\rbrace}

\newcommand{\p}{\partial}

\newcommand{\be}{\begin{equation}}
\newcommand{\bear}{\begin{eqnarray}}
\newcommand{\ear}{\end{eqnarray}}
\newcommand{\ee}{\end{equation}}
\newcommand{\lbl}{\label}
\newcommand{\bi}{\bibitem}
\newcommand{\ci}{\cite}

\newcommand{\vs}{\vspace}
\newcommand{\hs}{\hspace}

\begin{document}

\begin{center}

\


\baselineskip .7cm

{\bf \LARGE Point Particle with Extrinsic Curvature\\ as a Boundary
of a Nambu-Goto String:\\ Classical and Quantum Model}

\vs{2mm}

\baselineskip .5cm
Matej Pav\v si\v c

Jo\v zef Stefan Institute, Jamova 39,
1000 Ljubljana, Slovenia

e-mail: matej.pavsic@ijs.si

\vs{3mm}

{\bf Abstract}
\end{center}

\baselineskip .43cm

{\small It is shown how a string living in a higher dimensional
space can be approximated as a point particle with squared extrinsic curvature.
We consider a generalized Howe-Tucker action for such a ``rigid particle" and
consider its classical equations of motion and constraints. We find that
the algebra of the Dirac brackets between the dynamical variables associated
with velocity and acceleration contains the spin tensor. After quantization,
the corresponding operators can be represented by the Dirac matrices, projected
onto the hypersurface that is orthogonal to the direction of momentum. 
A condition for the consistency of such a representation is that the states
must satisfy the Dirac equation with a suitable effective mass. The
Pauli-Lubanski vector composed with such projected Dirac matrices is equal to
the Pauli-Lubanski vector composed with the usual, non projected, Dirac
matrices, and its eigenvalues thus correspond to spin one half states.
}

\vs{3mm}

\baselineskip .55cm

\section{Introduction}

Extended objects, such as branes with extrinsic curvature are of great
interest for physics\,\ci{Polyakov}--\ci{Pavsic1}.
A particular case is the point particle
with extrinsic curvature, the so called
 ``rigid particle''\ci{Nesterenko1}--\ci{Pavsic2b}.
 Such an object,
because of the second derivatives in the action, moves along a trajectory that
is not a straight line, but a helix. The rectilinear component of the helical
worldline corresponds to the particle's momentum $p_\mu$, whereas the
circular component is responsible for spin, $S_{\mu \nu}$. The quantities
$p_\mu$, $S_{\mu \nu}$ and the orbital momentum $L_{\mu \nu} =x_\mu p_\nu -
x_\nu p_\mu$ satisfy the relations of a classical particle with
spin\,\ci{Corben}--\ci{Kosyakov}. A question arises as to whether the rigid
particle can be a classical
model for the quantum particle with spin, described by the Dirac equation.
In fact, there are two types of rigid particles: those with the extrinsic
curvature to the power one (type 1)\,\ci{Pisarski1}--\ci{Banerjee},
and those with the squared intrinsic
curvature (type 2)\,\ci{Dereli}--\ci{Pavsic2b},\ci{Lindstrom2}.

It was shown\,\ci{Lindstrom2} that if one starts from an
ordinary Nambu-Goto string (without extrinsic curvature) living in a space with
one extra space-like dimension, then one can derive type 2 rigid particle
as an approximation.  According to the authors of Ref.\,\ci{Lindstrom2},
such derivation was not quite consistent. In Ref.\,\ci{Pavsic2a}, it was shown
how  we can obtain the consistent rigid particle:
the squared extrinsic curvature with the correct sign in the rigid particle's
action comes from a string living in a spacetime with an extra time-like
dimension. Starting from the Nambu-Goto string action, one can directly arrive
at the type 2 rigid particle action\,\ci{Lindstrom2,Pavsic2a}.
In this paper I will show, following the previous
work\,\ci{Pavsic2a}, that if we start from the Polyakov form
of the string action, then as an intermediate approximation we obtain
 a {\it generalized Howe-Tucker action} that contains second order
derivative\footnote{ If in the latter action we perform a further
approximation, then we obtain the type 2 rigid particle action.}. We will study the
classical and quantum equations of such a generalized Howe-Tucker point
particle action, describing what we will call type 2a rigid particle.
The system contains two first class and four second class
constraints\,\ci{Dirac}. The Dirac brackets between the phase
space variables associated
with velocity contains the spin tensor. The similar holds for the Dirac
brackets associated with acceleration. In the quantized theory, those Dirac
bracket relations become the commutation relations between
the operators\,\ci{Dirac}.
It turns out that these operators can be represented in terms of
the gamma matrices, multiplied by the generators of the Clifford algebra
$Cl(0,2)$ of a 2-dimensional space with signature $(--)$. The latter space
is a subspace of the phase space of our dynamical system. The signature
$(--)$ comes from the space like type of the chosen dynamical variables,
the 4-acceleration and the projection of the 4-velocity onto a space like
hypersurface. The Pauli-Lubanski vector turns out to be the same as that
for a Dirac particle. The analysis presented in this paper thus leads to
a conclusion that the type 2a rigid particle, upon quantization,
has spin $1/2$.
A. Deriglazov\,\ci{Deriglazov} considered type 1 rigid particle, whose
dynamics is different, but he also found that the phase space variables
can be quantized by gamma matrices and that the system has spin one-half.

The physical states must satisfy the conditions imposed by the first class
constraints. This can be consistent with the second class constraints and
the representation of the operators in terms of the Clifford numbers if we
bring an additional time-like dimension into the game, besides the two ones
considered so far in our model. 

In Sec.\,2 we describe a scenario with an open string living in a $(D+1)$-dimensional
target space whose $(D+1)^{\rm th}$ dimension, as well as the $1^{\rm th}$ one,
are time-like. For the $(D+1)^{\rm th}$ embedding function we
choose $X^{D+1} (\tau,\sigma)=\sigma$ which in the considered scenario,
illustrated in Fig.\,1,
is possible because of the reparametrization invariance of the string
action. We expand the string embedding functions $X^\mu (\tau,\sigma)$,
$\mu=0,1,2,...,D$. into the Taylor series around $\sigma=0$.  
If we take only two terms of the latter expansions, then
we obtain an action for a point particle which includes second order derivatives.
In Sec.\,3 we consider the dynamics of such a particle. We compute the
Hamiltonian, the corresponding equations of motion, the first and the second
class constraints and the Dirac brackets between certain dynamical variables.
In Sec.\,4 we perform quantization of our system in the Schr\"odinger and in
the Heisenberg picture. In Conclusion we summarize  our results and point
out why they are important for further progress on our road to the unified
theory of fields and particles, including quantum gravity. In Appendix we consider
a particular solution to the Laplace equation satisfied by the time-like string,
according to which the string end at $\sigma=0$ moves as a point particle with
extrinsic curvature.

\section{The string with time-like extension}

Let us consider an open string, embedded in a $(D+1)$-dimensional target
space, $M_{D+1}$, such that the string ends are attached to two $Dp$-branes\,\ci{Polchinski},
with $p=D-1$,  that sweep the worldvolumes $V_D$ and $V'_D$, as
shown in Fig.\,1. The extra, $(D+1)^{\rm th}$ dimension need not be compact,
it can extend to infinity. If so, we can adopt the braneworld
scenario\,(see, e.g., a review\,\ci{PavsicTapia}), and assume that our
world is in one of the two worldvolumes, say, in $V_{D}$. Since we observe
the 4-dimensional spacetime, we may take $D=4$, or assume that $D-4$
dimensions of $V_D$ are compactified. Alternatively, we can assume that
$V_D$ is the 16-dimensional Clifford space\,\ci{ExtendedClifford}, i.e., the space of
points, areas, volumes and 4-volumes associated with physical objects living
in 4-dimensional spacetime.
 We are interested in how the string
end moves in $V_D$. From the point of view of an observer in $V_{D}$, the motion of the string end is perceived as the motion of
a point particle.

\setlength{\unitlength}{.8mm}

\begin{figure}[h!]
\hs{3mm}
\begin{picture}(105,70)

\put(40,-10){\includegraphics[scale=1]{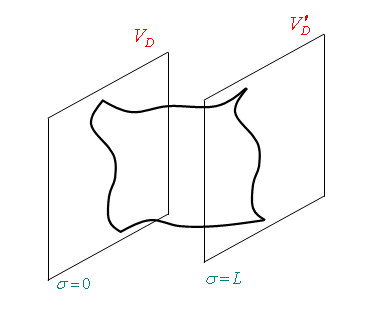}}

\end{picture}

\caption{\footnotesize A scenario in which a string's worldsheet is bounded by two
worldvolumes $V_D$ and $V'_D$, corresponding to two $Dp$-branes with $p=D-1$.}

\end{figure} 

The string is described by the Nambu-Goto action in
$(D+1)$-dimensions:
\be
    I = T \int \dd^2 \xi (\epsilon {\hat f})^{1/2},
\lbl{2.1}
\ee
where $T$ is the string tension, and
 ${\hat f} = {\rm det} \, {\hat f}_{a b}$, $~~{\hat f}_{a b} =
\p_a X^{\hat \mu} \p_b X_\hmu$, $~\hmu = (\mu,D+1)$, $\mu=0,1,2,...,D-1$,
$\xi^a=(\xi^1,\xi^2)\equiv (\tau,\sigma)$. Here $\epsilon = \pm 1$ depends on
the signature of the metric on the worldsheet $V_2$ swept by the string:
$\epsilon =-1$ corresponds to the signature $(+-)$, whereas $\epsilon =1$
corresponds to the signature $(++)$. Because the action (\ref{2.1}) is
invariant under reparametrizations of $\tau$ and $\sigma$, we have a
certain freedom in the choice of the string embedding functions
$X^\hmu (\tau,\sigma)$. In the scenario, illustrated in Fig.\,1, this enables us to set
\be
   X^{D+1} (\tau,\sigma) = \sigma
\lbl{2.2}
\ee

Let us now expand the remaining embedding functions $X^\mu (\tau,\sigma)$,
$\mu = 0,1,2,...,D-1$, into the Taylor series around $\sigma=0$:
\be
   X^\mu (\tau,\sigma) = X^\mu (\tau,0) 
   +\frac{\p X^\mu}{\p \sigma}\Biggl\vert_0 \sigma
   + \frac{1}{2} \frac{\p^2 X^\mu}{\p \sigma^2}\Biggl\vert_0 \sigma^2 + ...
\lbl{2.2a}
\ee

Introducing
\be
  x^\mu (\tau) \equiv X^\mu (\tau,0)~,
  ~~~y_1^\mu (\tau) \equiv \frac{1}{k} \frac{\p X^\mu}{\p \sigma}\Biggl\vert_0~,
  ~~~~y_2^\mu (\tau) \equiv \frac{1}{k^2} 
  \frac{\p^2 X^\mu}{\p \sigma^2}\Biggl\vert_0
\lbl{2.3}
\ee
the expansion (\ref{2.2a}) reads
\be
  X^\mu (\tau,\sigma) = x^\mu (\tau) + y_1^\mu (\tau) k \sigma + 
  \frac{1}{2} y_2^\mu (\tau) k^2 \sigma^2 + ...~,
\lbl{2.4}
\ee
where $k$ is a constant, such that the product $k \sigma$ is dimensionless.
We see that the string can be described in terms of infinite number
of the $\tau$-dependent functions $x^\mu (\tau)$, $y_i^\mu (\tau)$,
$i=1,2,..., \infty$. If the string is short, so that $k L \ll 1$,
already first few terms in the expansion (\ref{2.4}) will provide,
within a prescribed accuracy, a sufficiently good description of the
string, satisfying exactly the boundary conditions on one end, and
approximately on the other end.

The functions $X^\mu (\tau,\sigma)$, expanded according to Eq.\,(\ref{2.4}),
can be inserted into the action (\ref{2.1}). After performing the integration
over $\sigma$ from $0$ to $L$, we obtain an action  functional for an
infinite set of $\tau$-dependent variables $x^\mu (\tau)$, $y_i^\mu (\tau)$,
$i=1,2,..., \infty$. If we take
a finite number of terms in the expansion (\ref{2.4}), then we obtain an
action which is a functional of a finite number of variables
$x^\mu (\tau)$, $y_i^\mu (\tau)$, $i=1,2,...,n$. Our string with infinitely
many degrees of freedom is thus sampled by a finite number of degrees
of freedom. Descriptions of physical
objects in terms of an effective action are common in physics. Instead of
taking into account all the degrees of freedom, we can sample the object by less
degrees of freedom. For instance, though the Earth is an object with very
many degrees of freedom, we can describe its motion around the Sun,
by taking into account only its center of mass degrees of freedom.
We do not care about other variables and boundary conditions determining
their motion. Similarly, effective actions in which certain degrees of freedom
of a system have been integrated out, are commonly used in high energy
physics. According to (\ref{2.4}), a string can also be described by
an effective action, and the corresponding ``effective" equations of motion
for a finite set of
the variables $x^\mu (\tau)$, $y_i^\mu (\tau)$. Then, instead of the
boundary and initial conditions for the variables $X^\mu (\tau,\sigma)$,
we have to specify the initial conditions for $x^\mu (\tau)$, $y_i^\mu (\tau)$
only. Since we are interested only in the
behavior of a finite number of the variables $x^\mu (\tau)$, $y_i^\mu (\tau)$,
describing the motion of the string end at $\sigma=0$,
and not in the behavior of the entire string, we do not need to take into
account the boundary conditions for the string end at $\sigma=L$.

Let us now denote $y^\mu (\tau) \equiv y_1^\mu (\tau)$, write the
expansion (\ref{2.4}) as
\bear
    &&X^\mu (\tau,\sigma) = x^\mu (\tau) + y^\mu (\tau) k \sigma + ... \nonumber \\
   &&X^{D+1} (\sigma) = \sigma,
\lbl{2.6}
\ear
and neglect all higher order terms. Then, up to such
 approximation, the induced metric is
\be
     f_{ab}= \p_a X^\hmu \p_b X_\hmu=
        \begin{pmatrix} {\dot x}^2 + 2 {\dot x}{\dot y} k \sigma,  &
      {\dot x} y k\\
                 {\dot x} y k ,      &   k^2 y^2 + \epsilon \\
      \end{pmatrix} + ...\,,
\lbl{2.6a}
\ee
and the string action (\ref{2.1}) becomes\,\ci{Lindstrom2,Pavsic2a}
\be
   I = T \int_{0}^L \dd \tau \, \dd \sigma \, \sqrt{{\dot x}^2} \left (
   1 + \frac{\epsilon}{2} k^2 y^2 + \frac{{\dot x}{\dot y} k \sigma}{{\dot x}^2}
    - \frac{\epsilon k^2 ({\dot x} y)^2}{2 {\dot x}^2} \right )
     +  {\cal O} (k^2 L^2) ,
\lbl{2.7}
\ee
which, after the integration over $\sigma$, becomes
\be
   I = TL \int \dd \tau \,  \sqrt{{\dot x}^2} \left (
   1 + \frac{\epsilon}{2} k^2 y^2 + \frac{{\dot x}{\dot y} k L}{2{\dot x}^2}
    - \frac{\epsilon k^2 ({\dot x} y)^2}{2 {\dot x}^2} \right )
     +  {\cal O} (k^2 L^2) ,
\lbl{2.7a}
\ee
Variation of the latter action with respect to $y^\mu$ gives
\be
   y^\mu = - \frac{L}{2 k} H^\mu + \frac{1}{{\dot x}^2} ({\dot x}^\alpha
       y_\alpha) {\dot x}^\mu, ~~~
       H^\mu \equiv \frac{1}{\sqrt{{\dot x}^2}}\, \frac{\dd}{\dd \tau}
       \left ( \frac{{\dot x}^\mu}{\sqrt{{\dot x}^2}} \right ),
\lbl{2.8}
\ee
where $H^\mu {\dot x}_\mu = 0$. By inserting the expression (\ref{2.8}) into
the action (\ref{2.7}), and by writing $m=TL$ and $\mu=TL^3/8$, we obtain
\be
  I= \int \dd \tau \, \sqrt{{\dot x}^2} \, (m - \epsilon \mu H^2)
      + {\cal O} (k^2 L^2).
\lbl{2.9}
\ee
This is the action for the type 2 rigid particle, if $\epsilon=1$, i.e.,
if the worldsheet signature is $(++)$ and the $(D+1)$-th dimension is time-like.

Instead of the Nambu-Goto action (\ref{2.1}), we can consider the Polyakov
action
\be
     I[X^\hmu,\gam^{ab}] = \frac{T}{2} \int \dd^2 \xi \, \sqrt{\epsilon \gam}
     \, \gam^{ab} \p_a X^\hmu \p_b X_\hmu ,
\lbl{2.9a}
\ee
which gives the same classical equations of motion.

From now on, we will consider the case where $\epsilon=1$. Then
the signature of the target space is $(2,D-1)$, which
means two time-like and $D-1$ space-like dimensions. The direction
along $\sigma$ is bounded, whereas
the direction along $\tau$ is open. Until the seminal works by
Bars et all.\,\ci{Bars}, it was generally believed that in the presence
of extra time-like dimensions there must necessarily be ghosts that
cannot be eliminated, a consequence
being that such theories are inconsistent. However, the so called
2-time (2T) physics, developed in Refs.\,\ci{Bars}, is quite consistent
and gives numerous remarkable results. Moreover, in Refs.\,\ci{UltraHyperbolicOK}--\ci{PavsicBook}
it was shown that also the theories in ultra hyperbolic spaces, including
the self-interacting Pais-Uhlenbeck oscillators,
can be consistent and stable. Because of those encouraging results, it makes
sense to consider as well the target space with the signature $(2,D-1)$ in
which there lives a time-like string, sweeping a worldsheet with the signature
$(++)$.

By taking the expansion (\ref{2.6}) and inserting the metric (\ref{2.6a})
into the action (\ref{2.9a}), we obtain
\be
   I = \frac{T}{2} \int \dd \tau \, \dd \sigma \, \sqrt{\gam}
   \left [ \gam^{11} ({\dot x}^2 + 2 {\dot x} {\dot y} k \sigma) +
   2 \gam^{12} k {\dot x} y + \gam^{22}(k^2 y^2 + 1) \right ]
   + {\cal O} (k^2 L^2).
\lbl{2.10}
\ee
We now use $\gam^{ab} = (1/\gam)\p \gam/\p \gam_{ab}$, and
write
\be
   \sqrt{\gam} \gam^{11} = \frac{1}{\sqrt{\gam}} \gam_{22}
    = \frac{1}{E(\tau,\sigma)} = \frac{1}{e(\tau)} +
    \frac{\p E^{-1}}{\p \sigma}\Biggl\vert_0 \sigma + ...
\lbl{2.12}
\ee
\be
   -\sqrt{\gam} \gam^{12} = \frac{1}{\sqrt{\gam}} \gam_{12}
    = F(\tau,\sigma) = f(\tau) +
    \frac{\p F}{\p \sigma}\Biggl\vert_0 
    \lbl{2.13}
\ee
\be
   \sqrt{\gam} \gam^{22} = \frac{1}{\sqrt{\gam}} \gam_{11}
    = \frac{1}{E(\tau,\sigma)} = E (1+F^2)= e(1+f^2) + {\cal O}(\sigma)
\lbl{2.14}
\ee
Here $\gam= \gam_{11}\gam_{22} - \gam_{12}^2 = \gam [E(1+F^2)\frac{1}{E}-F^2]$,
which justifies the above parametrization.

If we insert Eqs.\,(\ref{2.12})--(\ref{2.14}) into the action (\ref{2.10})
and integrate over $\sigma$, we obtain
\be
     I = \frac{LT}{2} \int \dd \tau \, \left [ \frac{1}{e} ({\dot x}^2 + L k
   {\dot x}{\dot y}) + e (1+f^2) (k^2 y^2 + 1) - 2 f k {\dot x} y \right ]
   + {\cal O}(k^2 L^2) .
\lbl{2.15}
\ee
The equations of motion are:
\bear
   &&\delta e\; : ~~~~- \frac{1}{e^2} ({\dot x}^2 + L k{\dot x}{\dot y}) +
         (1+f^2) (1+ k^2 y^2) = 0, \lbl{2.16}\\
   &&\delta f\; : ~~~~ f e (1+k^2 y^2)- k {\dot x} y = 0, \lbl{2.17} \\
   && \delta y\; : - L k \, \frac{\dd}{\dd \tau} \left ( \frac{{\dot x}^\mu} {e} \right )
     + 2 e (1+f^2) y^\mu - 2 f k {\dot x}^\mu = 0,  \lbl{2.18}\\
   &&\delta x\; : ~~~~\frac{\dd}{\dd \tau} \left ( \frac{{\dot x}^\mu}{e}
   + \frac{L k {\dot y}^\mu}{2 e} - f k y^\mu \right ) = 0. \lbl{2.19}
\ear
Here $e$ and $f$ are the Lagrange multipliers. Choice of a Lagrange
multiplier means choice of gauge.  Recall that, according to
Eqs.\,(\ref{2.12})--(\ref{2.14}), $e$ and $f$ are related to the string metric.
The string action is invariant under reparametrizations of $\tau$ and $\sigma$,
a consequence being the existence of two constraints. By a judicious
choice of parameters $\tau$ and $\sigma$, one can transform the string metric
into a diagonal form (though not necessarily into the conformal one),
so that $\gamma^{12}$  vanish. But vanishing of $\gamma^{12}$ means
vanishing of $f$. 

If we choose $f=0$, then Eqs.\,(\ref{2.17}),(\ref{2.18}) give:
\be
    y^\mu {\dot x}_\mu = 0,
\lbl{2.20}
\ee
\be
    y^\mu = \frac{L}{2 k e}\,\frac{\dd}{\dd \tau} 
    \left ( \frac{{\dot x}^\mu} {e} \right ) = \frac{L}{2 k} \left (
    \frac{{\dot x}^2}{e^2}\,H^\mu + \frac{1}{e} \, 
    \frac{{\dot x}^\mu}{\sqrt{{\dot x}^2}}\, \frac{\dd}{\dd \tau}
    \left ( \frac{\sqrt{{\dot x}^2}}{e} \right ) \right ),
\lbl{2.20a}
\ee
where
\be
   H^\mu \equiv \frac{1}{\sqrt{{\dot x}^2}}\, \frac{\dd}{\dd \tau}
   \left ( \frac{{\dot x}^\mu}{\sqrt{{\dot x}^2}} \right ) .
\lbl{2.21}
\ee
From Eqs.\,(\ref{2.20})--(\ref{2.21}) we find
\be
  \frac{\dd}{\dd \tau} \left ( \frac{{\dot x}^\mu}{e} \right )
  \frac{{\dot x}^\mu}{\sqrt{{\dot x}^2}} = H^\mu {\dot x}_\mu 
  \frac{\sqrt{{\dot x}^2}}{e} + \frac{\dd}{\dd \tau} \left (
  \frac{\sqrt{{\dot x}^2}}{e} \right ) = 0.
\lbl{2.22}
\ee
By using the identity $H^\mu {\dot x}_\mu = 0$, we then obtain:
\be
   \frac{\dd}{\dd \tau} \left ( \frac{\sqrt{{\dot x}^2}}{e} \right ) = 0,
\lbl{2.23}
\ee
\be
   y^\mu = \frac{L}{2 k} \, \frac{{\dot x}^2}{e^2}\,H^\mu,
\lbl{2.24}
\ee
\be
   y^\mu y_\mu \equiv y^2 = \left ( \frac{L}{2 k} \right )^2
   \left ( \frac{{\dot x}^2}{e^2} \right )^2 H^2.
\lbl{2.25}
\ee
We see that $y^2$ is proportional to the squared extrinsic curvature
$H^2 \equiv H^\mu H_\mu$.

With $f=0$, the constraint (\ref{2.16}) reads
\be
  - \frac{1}{e^2} ({\dot x}^2 + L k {\dot x}{\dot y}) + 1 + k^2 y^2.
\lbl{2.26}
\ee
From Eqs.\,(\ref{2.20}),(\ref{2.20a}) it is not difficult to find
\be
   {\dot x}{\dot y} = - \frac{2 k}{L} e^2 y^2 .
\lbl{2.27}
\ee
Inserting this into Eq.\,(\ref{2.26}), we obtain:
\be
   \frac{{\dot x}^2}{e^2} = 1 + 3 k^2 y^2 .
\lbl{2.28}
\ee
Because ${\dot x}^2/{e^2}$ is a constant of motion (c.f. Eq.\,(\ref{2.23})),
also $y^2$ and $H^2$ are constants of motion.

From Eqs.\,(\ref{2.24}) and (\ref{2.28}) we have
\be
   \frac{{\dot x}^2}{e^2} = 1 + \frac{6 \mu}{m} 
   \left ( \frac{{\dot x}^2}{e^2} \right )^2 H^2 ,
\lbl{2.29}
\ee
where we have introduced
\be
   m= LT  ~~~~~ {\rm and} ~~~~\mu = \frac{L^3 T}{8} .
\lbl{2.30}
\ee
The solution of Eq.\,(\ref{2.29}) is
\be
    \frac{{\dot x}^2}{e^2}= \frac{1 \pm \sqrt{1-\frac{24 \mu}{m}H^2}}
    {\frac{12 \mu H^2}{m}} .
\lbl{2.31}
\ee
Assuming that $24 \mu H^2/{m} \ll 1$, we obtain:
\bear
    &&~{\rm I.}~~~~\frac{{\dot x}^2}{e^2} \approx \frac{m}{6 \mu H^2}
     \hs{2cm} , \lbl{2.32}\\
    &&{\rm II.}~~~~\frac{{\dot x}^2}{e^2} \approx 1 + \frac{6 \mu H^2}{m} 
    \hs{2cm}.
\lbl{2.33}
\ear
Solution I is inconsistent with our assumption that ${\dot x}^2>0$ and
$H^2 <0$. Therefore we take Solution II, and write
\be
   \frac{\sqrt{{\dot x}^2}}{e} \approx \sqrt{1+ \frac{6 \mu H^2}{m}}
   \approx 1+ \frac{3 \mu H^2}{m} .
\lbl{2.34}
\ee

By using Eqs.\,(\ref{2.27}),(\ref{2.34}),(\ref{2.30}) and $f=0$, the action
(\ref{2.15}) becomes
\be
   I = \int \dd \tau \, \sqrt{{\dot x}^2}\, (m - \mu H^2) + {\cal O}
       \left ( m \left ( \frac{\mu H^2}{m} \right )^2  \right ),
\lbl{2.35}
\ee
which is the {\it rigid particle action}.

We see that as a first approximation to the string action (\ref{2.9a}) we
obtain the point particle action (\ref{2.15}). A further approximation is
in expanding ${\dot x}^2/e^2$ according to (\ref{2.33}) and then
$\sqrt{{\dot x}^2}/e$ according to (\ref{2.34}). Then we obtain the type
2 rigid particle action, apart from the term $m (\mu H^2/m)^2$ and the
higher order terms that we neglect. In the rest of the paper we will
consider the ``intermediate", type 2a, action (\ref{2.15}), and its equivalent forms.

\section{The dynamics of the spinning point particle\\ derived from the string}

In the previous section we derived from the Polyakov string action (\ref{2.9a})
an effective point particle action which for $f=0$ reads
\be
   I = \frac{L T}{2} \int \dd \tau \, \left [ \frac{{\dot x}^2}{e}+e +
   \frac{Lk {\dot x}{\dot y}}{e} + e k^2 y^2 \right ].
\lbl{3.1}
\ee
Let me clarify again that
this action is an approximation to the Polyakov action in the sense that it does not
describe all the degrees of freedom of the string; it describes the motion of the string
end at $\sigma=0$, attached to the $Dp$=brane $V_{D-1}$.
Even less degrees of freedom we take into account if we consider the expression
$X^\mu (\tau,\sigma) = x^\mu(\tau)$, in which we neglect  all terms in the
expansion (\ref{2.6}), except the first one, i.e., if we take $k=0$.
Then, instead of the action (\ref{3.1}), we obtain the well known
Howe-Tucker\,\ci{HoweTucker} action for a point particle: it describes the
center of mass motion of the string. The boundary conditions for the string's
ends are irrelevant for its center of mass. Instead of the string equations
of motion we have then the equations of motion for its center of mass.
What we have to specify in such a case are the initial conditions for
the center of mass coordinates.
The action (\ref{3.1}), besides $x^\mu (\tau)$,  contains also
the $y^\mu (\tau)$ degrees of freedom that are related to the extrinsic
curvature, making the twist along a helix, as illustrated in Fig. 2
(see Appendix). The motion of the string end at $\sigma=L$ is irrelevant
for the degrees of freedom $x(\tau)$ and $y(\tau)$, describing, respectively,
the motion of the string's end at $\sigma=0$, and the twist along a helix.
Notice that the parameter $\sigma$ does not occur at all in the degrees of
freedom $x$ and $y$.
Having in mind this fact, it is obvious, why in the quenched description of
the string in terms of $x$ and $y$ it is not necessary
to take into account the string's boundary conditions. Boundary conditions
are relevant for those degrees of freedom that depend on $\sigma$. Here,
the degrees of freedom $x^\mu$ and $y^\mu$ depend on $\tau$ only, therefore
their evolution in $\tau$ can be determined from the equations of motion if one
provides initial conditions, e.g., at $\tau=0$.

If we plug the solution
\be
   y^\mu = \frac{L}{2 k} \,\frac{1}{e}\, \frac{\dd}{\dd \tau}
           \left ( \frac{{\dot x}^\mu}{e} \right )
\lbl{3.2}
\ee
into the action (\ref{3.1}) and introduce parameters $m$ and $\mu$ according
to (\ref{2.30}), then we obtain\footnote{
Introducing a new parameter $\tau'$ according to
$\dd \tau' m {\tl e} = \dd \tau e$, the action (\ref{3.3}) assumes the form
$I[x^\mu,{\tl e}] = \int \dd \tau \, \left [ \frac{1}{2} \left ( 
  \frac{{\dot x}^2}{\tl e}+{\tl e} m^2 \right ) 
  - \frac{\mu}{m^3 {\tl e}} \, \frac{\dd}{\dd \tau}
  \left (\frac{{\dot x}^\mu}{\tl e} \right ) \frac{\dd}{\dd \tau}
  \left (\frac{{\dot x}_\mu}{\tl e} \right ) \right ]$,
where we have renamed $\tau'$ into $\tau$.
From Eq.\,(\ref{2.30}) we have $\mu/m^3 = 1/(8T^2) \equiv {\tl \mu}$.
In the limit $m \rightarrow 0$, such that $L \rightarrow 0$, $T = finite$,
the latter action becomes identical to
the  action for the ''massless" particle with curvature, considered
by McKeon\,\ci{McKeon}.}
\be
   I[x^\mu,e] = \int \dd \tau \, \left [ \frac{m}{2} \left ( 
  \frac{{\dot x}^2}{e}+e \right ) - \frac{\mu}{e} \, \frac{\dd}{\dd \tau}
  \left (\frac{{\dot x}^\mu}{e} \right ) \frac{\dd}{\dd \tau}
  \left (\frac{{\dot x}_\mu}{e} \right ) \right ] .
\lbl{3.3}
\ee

The action (\ref{3.3}) contains second order derivative. We will now employ the
Ostrogradsky formalism\,\ci{Ostrogradski} and transform (\ref{3.3}) into
the Hamilton form. The momenta are
\bear
  &&p_\mu = \frac{\p L}{\p {\dot x}^\mu} 
  - \ddt \left (\frac{\p L}{\p {\ddot x}^\mu} \right )
  = \frac{m {\dot x}_\mu}{e} + \frac{2 \mu}{e}\, \ddt \left ( \frac{1}{e}\,
  \ddt \left ( \frac{{\dot x}_\mu}{e} \right ) \right ) , \lbl{3.4} \\
  &&\pi_\mu = \frac{\p L}{\p {\ddot x}^\mu} = - \frac{2 \mu}{e^2} \,
  \ddt \left ( \frac{{\dot x}_\mu}{e} \right ) , \lbl{3.5} \\
  &&p_e = \frac{\p L}{{\p \dot e}} = \frac{2 \mu}{e^3}\, {\dot x}^\mu \, \ddt
  \left ( \frac{{\dot x}_\mu}{e} \right ) .   \lbl{3.6}
\ear
The equation of motion for $x^\mu$ is:
\be
   \delta x^\mu : ~~~~~~ {\dot p}_\mu = 0 \hs{1.6cm}.
\lbl{3.7}
\ee
The equation of motion for $e$ is
\be
    \delta e : ~~~~~\frac{\p L}{\p e} - \ddt \frac{\p L}{\p{\dot e}} = 0,
\lbl{3.8}
\ee
which gives
\be
   \frac{m}{2} \left ( 1 - \frac{{\dot x}^2}{e^2} \right ) + 3 \mu \,\frac{1}{e}
   \ddt \left ( \frac{{\dot x}^\mu}{e} \right )\frac{1}{e}
   \ddt \left ( \frac{{\dot x}_\mu}{e} \right ) - \frac{2 \mu}{e}\,
   \ddt \left ( \frac{{\dot x}^\mu}{e^2} 
   \ddt \left ( \frac{{\dot x}_\mu}{e} \right )  \right ) = 0 .
\lbl{3.9}
\ee
The Hamiltonian is
\be
   H_0 = p_\mu \dotx^\mu + \pi_\mu {\ddot x}^\mu + p_e {\dot e} - L_0 ,
\lbl{3.10}
\ee
where $L_0$ is the Lagrangian corresponding to the action (\ref{3.3}).
By introducing
\be
   \dotx^\mu = q^\mu,~~~~~~{\dot e} = \beta ,
\lbl{3.10a}
\ee
and by inserting (see Eqs.\,(\ref{3.5}) and (\ref{3.6}))
\bear
   &&{\ddot x}^\mu = - \frac{e^3}{2 \mu} \pi^\mu + \frac{{\dot e}}{e} q^\mu,
   \lbl{3.11}\\
   &&p_e = - \frac{\pi_\mu q^\mu}{e} \lbl{3.12}
\ear
into Eq.\,(\ref{3.10}), we obtain
\be
   H_0 = p_\mu q^\mu - \frac{e^3 \pi^2}{4 \mu} - \frac{m}{2}
   \left ( \frac{q^2}{e} +e \right ) + \beta ( p_e + \frac{\pi_\mu q^\mu}{e} ) ,
\lbl{3.13}
\ee
and
\be
  L_0 = p_\mu \dotx^\mu + \pi_\mu {\dot q}^\mu + p_e {\dot e} -
  \left [p_\mu q^\mu - \frac{e^3 \pi^2}{4 \mu} - \frac{m}{2}
   \left ( \frac{q^2}{e} +e \right ) \right ] 
   - \beta  ( p_e + \frac{\pi_\mu q^\mu}{e} ) .
\lbl{3.14}
\ee
This Lagrangian gives the equations of motion which are equivalent to
Eqs.\,(\ref{3.4})--(\ref{3.8}).

Let us now take into account the fact that the action (\ref{3.1}) holds
for $f=0$ which, according to Eq.\,(\ref{2.17}), implies
\be
    \dotx^\mu y_\mu = 0.
\lbl{3.15}
\ee
In view of Eqs.\,(\ref{3.2}),(\ref{3.5}) and (\ref{3.10a}), the above
relation reads
\be
   \pi_\mu q^\mu = 0 .
\lbl{3.16}
\ee
By imposing the latter constraint, the Lagrangian (\ref{3.14}) is
extended to
\be
  L =  p_\mu \dotx^\mu + \pi_\mu {\dot q}^\mu + p_e {\dot e} -
  \left [p_\mu q^\mu - \frac{e^3 \pi^2}{4 \mu} - \frac{m}{2}
   \left ( \frac{q^2}{e} +e \right ) \right ] - \alpha \pi_\mu q^\mu
   - \beta  ( p_e + \frac{\pi_\mu q^\mu}{e} ) .
\lbl{3.17}
\ee
The corresponding Hamiltonian is
\be
   H = p_\mu q^\mu - \frac{e^3 \pi^2}{4 \mu} - \frac{m}{2}
   \left ( \frac{q^2}{e} +e \right ) +\alpha \pi_\mu q^\mu
   + \beta  ( p_e + \frac{\pi_\mu q^\mu}{e} ) .
\lbl{3.18}
\ee
It gives the following equations of motion:
\bear
  &&\dotx^\mu =\lbrace x^\mu, H \rbrace = q^\mu, \lbl{3.18a}\\
  &&{\dot e} = \lbrace e, H \rbrace = \beta , \lbl{3.19}\\
  &&{\dot q}^\mu = \lbrace q^\mu,H \rbrace = - \frac{e^3 \pi^\mu}{2 \mu}
  + \alpha q^\mu + \frac{\beta q^\mu}{e} , \lbl{3.20}\\
  &&{\dot p}_\mu = \lbrace p_\mu , H \rbrace = 0 ,  \lbl{3.21}\\
  &&{\dot \pi}_\mu = \lbrace \pi_\mu, H \rbrace =
   - \left ( p_\mu - \frac{m q_\mu}{e} + \alpha \pi_\mu
    + \frac{\beta \pi_\mu}{e} \right ) , \lbl{3.22}\\
  &&{\dot p}_e = \lbrace p_e , H \rbrace = - \frac{3 e^2 \pi^2}{4 \mu}
  - \frac{m}{2} \left ( 1 - \frac{q^2}{e^2} \right ) - \beta
  \frac{\pi_\mu q^\mu}{e^2} , \lbl{3.23}
\ear
Using (\ref{3.12}), (\ref{3.16}), and (\ref{3.23}), we find that our
system, described by the Hamiltonian (\ref{3.18}), has the following
constraints:\footnote{The above equations and the analogous equations throughout this paper
are valid on the constraint surface $\Sigma$, but for
simplicity we write the equality symbol``\,=\," instead of 
a weak equality symbol such as ``\, $\stackrel{\Sigma}{=}$\,''.}
\bear
  &&{\tl \phi}_1 = \frac{3 e^2 \pi^2}{4 \mu} 
  + \frac{m}{2} \left ( 1 - \frac{q^2}{e^2} \right ) = 0 , \lbl{3.24}\\
  &&\boxed{\phi_2 = \pi_\mu q^\mu = 0} , \lbl{3.25}
\ear
where $q^2 \equiv q^\mu q_\mu$ and $\pi^2 \equiv \pi^\mu \pi_\mu$.

From the requirement that those constraints must be conserved, we obtain
further constraints:
\bear
  &&{\dot {\tl \phi}}_1 = 0  ~~~ \Rightarrow ~~~- \frac{3 e^2}{2 \mu} (\pi_\mu p^\mu
  + \alpha \pi^2) - \alpha \frac{m q^2}{e^2} = 0 , \lbl{3.27}\\
  &&{\dot \phi}_2 = 0 ~~~ \Rightarrow ~~~\boxed{\phi_4 = \frac{p_\mu q^\mu}{e}
 +  \frac{e^2 \pi^2}{2 \mu}- \frac{m q^2}{e^2} = 0} , \lbl{3.28}\\
  &&{\dot \phi}_4 = 0 ~~~\Rightarrow ~~~ - \frac{3 e^2 p_\mu \pi^\mu}{2 \mu}
   + \alpha \left ( \frac{p_\mu q^\mu}{e} - \frac{2 m q^2}{e^2}
   - \frac{e^2 \pi^2}{\mu} \right ) = 0 . \lbl{3.29}
\ear
If we substract Eq.\,(\ref{3.27}) from Eq.\,(\ref{3.29}), we obtain
$\alpha \phi_4 = 0$. Let us introduce the linear combination 
\be
  \boxed{{\phi}_1 = - {\tl \phi}_1 + \phi_4 = \frac{p_\mu q^\mu}{e} -
  \frac{m}{2} \left ( 1 + \frac{q^2}{e^2} \right ) - \frac{e^2 \pi^2}{4 \mu}} .
\lbl{3.30}
\ee
The conservation of $\phi_1$ gives
\bear
   {\dot \phi}_1 = 0 ~~~ &&\Rightarrow ~~~ - \frac{e^2}{\mu} p_\mu \pi^\mu
   + \alpha \phi_4 = 0, \nonumber \\
   &&{\rm i.e.} ~~~~ \boxed{\phi_3 = e p_\mu \pi^\mu = 0}.
\lbl{3.31}
\ear
From Eqs.\,(\ref{3.31}),(\ref{3.24}) and (\ref{3.27}) it follows that
$\alpha = 0$. The other Lagrange multiplier, $\beta$, has already been
determined by Eq.\,(\ref{3.19}).

Further we have
\bear
  &&{\dot \phi}_3 = 0 ~~~ \Rightarrow ~~~ 
  \boxed{\phi_5 = p^2 - \frac{m p_\mu q^\mu}{e}= 0} , \lbl{3.32} \\
  &&{\dot \phi}_5 = 0 ~~~ \Rightarrow ~~~ \frac{m e^2}{2 \mu} p_\mu \pi^\mu 
  - \alpha \frac{m p_\mu q^\mu}{e} = 0 . \lbl{3.33}
\ear
Because $\alpha = 0$, the last equation gives $\phi_3 = 0$, which is not a new
constraint.

Finaly, from Eq.\,(\ref{3.12}) we obtain the constraint
\be
  \boxed{\phi_6 = e p_e + \pi_\mu q^\mu = 0} .
\lbl{3.26}
\ee
 The conservation of the latter constraint, i.e., 
 ${\dot \phi}_6 = \{\phi_6,H\}=0$, does not give a new constraint.
 
We have thus the six constraints $\phi_\alpha$, $\alpha = 1,2,3,4,5,6$.
The reason why we have introduced the linear combination
$\phi_1 = - {\tl \phi}_1 + \phi_4$ (see Eq.\,(\ref{3.30})), is in the fact 
that its Poisson brackets with all the remaining constraints vanish.
The same holds for $\phi_6$. Therefore, $\phi_1$ and $\phi_6$ are
{\it first class constraints}, whereas $\phi_{\bar \alpha}$, ${\bar \alpha}
= 2,3,4,5$, are {\it second class constraints}.

In summary, we have {\it two first class constraints}
\bear
 &&\phi_1 = \frac{p_\mu q^\mu}{e} -
  \frac{m}{2} \left ( 1 + \frac{q^2}{e^2} \right ) - \frac{e^2 \pi^2}{4 \mu}
  \lbl{vez1}\\
  &&\phi_6 = e p_e + \pi_\mu q^\mu \lbl{vez6}
\ear
and {\it four second class constraints}
\bear
  &&\phi_2 = \pi_\mu q^\mu \lbl{phi2}\\
  && \phi_3 = e p_\mu \pi^\mu \lbl{phi3}\\
  &&\phi_4 = \frac{p_\mu q^\mu}{e}
    +\frac{e^2 \pi^2}{2 \mu}- \frac{m q^2}{e^2} \lbl{phi4}\\
  &&\phi_5 = p^2 - \frac{m p_\mu q^\mu}{e} \lbl{phi5} .
\ear
The presence of two first class constraints is a result of the fact that we derived our
rigid particle from the string, which has two gauge degrees of freedom, related to the
worldsheet parameters $\tau$ and $\sigma$. A reparametrization of those two parameters  
induces a change of $x^\mu$ and $y^\mu$ in the expansion (\ref{2.6}), which is reflected in a change of
our dynamical variables, such a change being generated by the first class constraints $\phi_1$ and $\phi_6$.

If we write the total Hamiltonian,  $H_{\rm tot} = H + \lambda_3 \phi_3
+ \lambda_4 \phi_4 + \lambda_5 \phi_5$, where $H$ is given in Eq.\,(\ref{3.18}),
and calculate
\be
   {\dot \phi}_i = \lbrace \phi_i , H_{\rm tot} \rbrace, ~~~i = 3,4,5
\lbl{3.34}
\ee
we find $\lambda_3=\lambda_4 = \lambda_5 = 0$.  According to the definition (\ref{3.18}), 
$H$ is a superposition of $\phi_1$, $\phi_2$ and $\phi_6$, the corresponding Lagrange
multipliers being $\lambda_1=1$, $\lambda_2 = \alpha =0$, and $\lambda _6 = \beta = {\dot e}$,
respectively.  All Lagrange multipliers for the second class constraints thus vanish. The Lagrange
multipliers for the first class constraints can be arbitrary, they need not be fixed to $\lambda_1=1$
and $\lambda_6 = {\dot e}$. In general, the total Hamiltonian is thus $H_{\rm tot} = \lambda_1 \phi_1
+ \lambda_6 \phi_6$, with arbitrary $\lambda_1$, $\lambda_6$.

We are now interested in the behaviour of certain quantities on the constraint surface $\Sigma$.
From the system of equations\footnote{Here $\phi_1=0$ does not mean that $\phi_1$ is strongly
zero, it is zero on the constraint surface only  (see footnote 3).}
  $\phi_1=0$, $\phi_4 =0$, $\phi_5=0$,  valid on $\Sigma$, we can
calculate $p_\mu q^\mu/e$, $m q^2/e^2$  and $ e^2 \pi^2/\mu$ in terms
of $p^2 \equiv p^\mu p_\mu = M^2$, where $M^2$ is a constant of motion.
We obtain:
\bear
    && \frac{p_\mu q^\mu}{e} = \frac{M^2}{m}, \lbl{3.35}\\
    && \frac{m q^2}{e^2} =\frac{m}{2} \left ( \frac{3 M^2}{m^2}- 1 \right ),
    \lbl{3.36}\\
    && \frac{e^2 \pi^2}{\mu} = m \left (\frac{M^2}{m^2} - 1 \right ). \lbl{3.37}
\ear
Because $q^\mu = {\dot x}^\mu$ is a time-like vector, we have
\be
    0 \le \frac{q^2}{e^2} \le 1 .
\lbl{3.38}
\ee
On the other hand, $\pi_\mu$ is proportional to the acceleration,
therefore we have
\be
    \pi^2 \equiv \pi^\mu \pi_\mu \le 0.
\lbl{3.39}
\ee
The condition that (\ref{3.38}) and (\ref{3.39}) are simultaneously
satisfied  is
\be
    \frac{1}{3} \le \frac{M^2}{m^2} < 1 .
\lbl{3.40}
\ee

The Poisson brackets of the {\it first class constraints} $\phi_1$ and $\phi_6$
with all the constraints vanish. The Poisson brackets between the remaining
constraints, $\phi_{\bar \alpha}$, ${\bar \alpha} = 2,3,4,5$, do not all vanish,
therefore these are  {\it second class constraints}. If we calculate the matrix
$C_{{\bar \alpha}{\bar \beta}} = \lbrace \phi_{\bar \alpha},\phi_{\bar \beta}
\rbrace$, we obtain on the constraint surface that
\be   
    C_{{\bar \alpha}{\bar \beta}} = 
    \begin{pmatrix}  
    0        &                           0   & \frac{3 M^2}{m}-2 m & M^2 \\
    0        &                           0   &        M^2       &    M^2 \\
  -\left (\frac{3 M^2}{m}-2 m \right ) & -M^2    & 0      &          0 \\
  -M^2       &                           -M^2    &  0      &          0
  \end{pmatrix}
\lbl{3.41}
\ee
Its determinat is  det $C_{{\bar \alpha}{\bar \beta}} = 4 M^4 (m^2-M^2)^2$.
The reciprocal matrix is
\be
   C^{{\bar \alpha}{\bar \beta}} = \frac{1}{2 (m^2-M^2)}
   \begin{pmatrix}
   0  & 0 & m & -1 \\
   0  & 0 & -1 & - \frac{2 m^2-3 M^2}{m M^2} \\
  -m  & 1 & 0 & 0 \\
   1 & \frac{2 m^2-3 M^2}{m M^2} & 0 & 0
   \end{pmatrix}
\lbl{3.42}
\ee

Because of the presence of the second class constraints it is convenient
to introduce the {\it Dirac brackets}, which are the projections of the
Poisson brackets onto the constraint surface:
\be
  \lbrace F,G \rbrace_D = \lbrace F,G \rbrace - \lbrace F, \phi_{\bar \alpha}
  \rbrace C^{{\bar \alpha}{\bar \beta}} \lbrace \phi_{\beta} , G \rbrace,
\lbl{3.43}
\ee
where $F$, $G$ are phase space functions.

In particular, we have\footnote{
By this notation we mean that $\lbrace q^\mu,\phi_2 \rbrace = q^\mu$,
$\lbrace q^\mu,\phi_3 \rbrace = e p^\mu$, etc.\,.
}
\be
   \lbrace q^\mu , \phi_{\bar \alpha} \rbrace = 
   \left (q^\mu, e p^\mu, \frac{e^2 \pi^\mu}{\mu}, 0 \right ) ,
\lbl{3.43a}
\ee
\be
   \lbrace \pi^\mu , \phi_{\bar \alpha} \rbrace
    = \left (-\pi^\mu, 0, - \frac{p^\mu}{e} +
   \frac{2 m q^\mu}{e^2}, \frac{m p^\mu}{e} \right ) , 
\lbl{3.43b}
\ee
from which we can calculate the following Dirac brackets:
\be
  \lbrace q^\mu , q^\nu \rbrace_D = \frac{e^2 m}{2 \mu (m^2-M^2)}
  \left ( S^{\mu \nu} + \frac{e}{m} \pi^{[\mu} p^{\nu]} \right ) ,
\lbl{3.44}
\ee
 \be
  \lbrace \pi^\mu , \pi^\nu \rbrace_D = \frac{m^2}{(m^2-M^2) e^2}
  \left ( S^{\mu \nu} + \frac{e}{m} \pi^{[\mu} p^{\nu]} \right ),
\lbl{3.44a}
\ee
\be
   \lbrace q^\mu , \pi^\nu \rbrace_D = \eta^{\mu \nu} +
   \frac{1}{2 (m^2-M^2)} \left [ \frac{e^2 m}{\mu}\pi^\mu \pi^\nu +
   2 \left ( 2 - \frac{m^2}{M^2} \right ) p^\mu p^\nu + \frac{2 m^2}{e^2} q^\mu q^\nu 
   - \frac{2 m}{e} (q^\mu p^\nu + q^\nu p^\mu) \right ]
\lbl{3.44b}
\ee
Here $S^{\mu \nu} = q^\mu \pi^\nu - q^\nu \pi^\mu$ is {\it the spin tensor}.
It is the orbital momentum in the $q^\mu$-space. Together with the
orbital momentum in the $x^\mu$-space it forms {\it the total angular
momentum} $J^{\mu \nu} = L^{\mu \nu} + S^{\mu \nu}$, which is a conserved
quantity for our action associated with the
Lagrangian (\ref{3.17})\,\ci{Riewe,Plyushchay,Pavsic1}.

The extra term $(e/m) \pi^{[\mu} p^{\nu]} $ modifies the spin tensor
$S^{\mu \nu}$ into
\be
  {\tl S}^{\mu \nu} = V^\mu \pi^\nu - V^\nu \pi^\mu ,
\lbl{3.46}
\ee
where
\be
   V^\mu = q^\mu - \frac{e p^\mu}{m} ,
\lbl{3.46aa}
\ee
so that we have
\be
  \lbrace q^\mu,q^\nu \rbrace_D = \lbrace V^\mu , V^\nu \rbrace_D 
   = \frac{e^2 m}{2 \mu (m^2-M^2)} {\tl S}^{\mu \nu}
\lbl{3.46a}
\ee
\be
  \lbrace \pi^\mu,\pi^\nu \rbrace_D 
  = \frac{m^2}{e^2 (m^2-M^2)} {\tl S}^{\mu \nu}
\lbl{3.46b}
\ee
The spin tensor so modified is in fact the spin tensor subjected to the
translation in the $q^\mu$-space, according to Eq.\,(\ref{3.46aa}). In other
words, it is the $q^\mu$-space orbital momentum
translated by the vectors $(e/m) p^\mu$. From the equations of motion
it is straightforward to derive that ${\tl S}^{\mu \nu}$ is a constant
of motion, $\dd{\tl S}^{\mu \nu}/\dd \tau=0$, whereas
$\dd S^{\mu \nu}/\dd \tau=p^\mu q^\nu-p^\nu q^\mu$.
The Pauli-Lubanski pseudovector
is the same for $S^{\mu \nu}$ and ${\tl S}^{\mu \nu}$:
\be
   S_\mu = \frac{1}{2M} \epsilon_{\mu \nu \rho \sigma} J^{\nu \rho} p^\sigma
  = \frac{1}{2 M} \epsilon_{\mu \nu \rho \sigma} S^{\nu \rho} p^\sigma
  =\frac{1}{2 M} \epsilon_{\mu \nu \rho \sigma} {\tl S}^{\nu \rho} p^\sigma ,
\lbl{3.47}
\ee
and it is a constant of motion.

By using (\ref{3.46aa}) we will now simplify Eq.\,(\ref{3.44b}) as well.
Since
\be
  V^\mu V^\nu = \left ( q^\mu - \frac{e p^\mu}{m} \right )
  \left ( q^\nu - \frac{e p^\nu}{m} \right ) =
  q^\mu q^\nu - \frac{e}{m} (q^\mu p^\nu +q^\nu p^\mu) 
  + \frac{e^2}{m^2} p^\mu p^\nu
\lbl{3.47a}
\ee
and
\be
  2 \left ( 2 - \frac{m^2}{M^2} \right ) =
  2 \left ( 1 - \frac{m^2}{M^2} \right ) + 2,
\lbl{3.47b}
\ee
we obtain
\be
  \lbrace q^\mu, \pi^\nu \rbrace_D 
  = \lbrace V^\mu, \pi^\nu \rbrace_D =\eta^{\mu \nu} - \frac{p^\mu p^\nu}{M^2}
  + \frac{1}{2 (m^2-M^2)} \left [ \frac{e^2 m^2 }{\mu} \pi^\mu \pi^\nu
  + \frac{2 m^2}{e^2} V^\mu V^\nu \right ].
\lbl{3.48}
\ee

Let us also calculate the following Dirac bracket:
\be
  \lbrace q^\mu, S^{\mu \nu} \rbrace_D = \lbrace q^\mu, S^{\mu \nu} \rbrace +
   \lbrace q^\rho , \phi_{\bar \alpha} \rbrace  C^{{\bar \alpha}{\bar \beta}}
   \lbrace S^{\mu \nu}, \phi{\bar \beta} \rbrace .
\lbl{3.49}
\ee
Using the relations,
\be
  \lbrace S^{\mu \nu}, \phi_{\bar \alpha} \rbrace =
  \left ( 0, e p^{[\mu}\pi^{\nu]}, \frac{1}{e} p^{[\mu} q^{\nu]},
  - \frac{m}{e} p^{[\mu} q^{\nu]} \right ) ,
\lbl{3.49a}
\ee
we obtain
\be
  \lbrace q^\rho, S^{\mu \nu} \rbrace_D = q^\mu \eta^{\nu \rho} -
  q^\nu {\eta}^{\mu \rho} + \frac{1}{2 (m^2-M^2)} \left [ \frac{e^3}{\mu}
  \pi^\rho p^{[\mu}\pi^{\nu]} +  \frac{2 (m^2-2M^2)}{M^2} 
  p^\rho \, p^{[\mu} q^{\nu]} 
  +\frac{2m}{e} q^\rho p^{[\mu} p^{\nu]} \right ] ,
\lbl{3.50}
\ee
which is a rather complicated expression. Let us now calculate
\be
   \lbrace V^\rho , {\tl S}^{\mu \nu} \rbrace_D 
   = \lbrace q^\rho , S^{\mu \nu} \rbrace_D + \lbrace q^\rho , \frac{e}{m}
     \pi^{[\mu}p^{\nu]} \rbrace_D
\lbl{3.51}
\ee
Using Eqs.\,(\ref{3.42}),(\ref{3.43a}) and
\be
   \lbrace \pi^{[\mu}p^{\nu]}, \phi_{\bar \alpha} \rbrace =
   \left ( - \pi^{[\mu}p^{\nu]},0, \frac{2 m}{e^2} q^{[\mu} p^{\nu]},0 \right ) ,
\lbl{3.52}
\ee
we obtain
\bear
   \lbrace V^\rho , {\tl S}^{\mu \nu} \rbrace_D &=& V^\mu \eta^{\nu \rho}
   - V^\nu \eta^{\mu \rho} + \frac{1}{2 (m^2-M^2)} \left ( 1 +
   \frac{2m^2-3M^2}{M^2} \right ) p^\rho p^{[\mu} q^{\nu]} \nonumber\\
   &=& V^\mu \eta^{\nu \rho} - V^\nu \eta^{\mu \rho} 
   + \frac{1}{M^2} p^\rho \, p^{[\mu} q^{\nu]} \nonumber\\
   &=& V^\mu \left (\eta^{\nu \rho} - \frac{1}{M^2} p^\nu p^\rho \right ) 
    - V^\nu \left (\eta^{\mu \rho} - \frac{1}{M^2} p^\mu p^\rho \right ) ,
\lbl{3.53}
\ear
where we have put $p^{[\mu} q^{\nu]} = p^{[\mu} (q^{\nu]} - (e/m) p^{\nu]}) =
p^{[\mu} V^{\nu]}$.

Similarly, we obtain
\be
  \lbrace \pi^\rho, {\tl S}^{\mu \nu} \rbrace_D =
  \pi^\mu \left (\eta^{\nu \rho} - \frac{1}{M^2} p^\nu p^\rho \right ) 
    - \pi^\nu \left (\eta^{\mu \rho} - \frac{1}{M^2} p^\mu p^\rho \right ) 
\lbl{3.54}
\ee

Using (\ref{3.46aa}), the constraint (\ref{3.32}) can be written in the
form
\be
     V^\mu p_\mu = 0.
\lbl{3.55}
\ee
The algebra of the Dirac brackets (\ref{3.46a}),(\ref{3.46b}),
(\ref{3.53}) and (\ref{3.54}) resembles
the algebra of the quantum commutators that are satisfied by the Dirac
matrices. A difference is in Eqs.\,(\ref{3.53}),(\ref{3.54}),
where the Minkowski metric
$\eta^{\nu \rho}$ is changed into the modified metric
$\eta^{\mu \rho} - p^\mu p^\rho/M^2$.

\vs{2mm}

{\bf {\it Equations of motion in terms of the Dirac brackets}}

\vs{1mm}

If in eqs.\,(\ref{3.18a})--(\ref{3.23}) we replace the Poisson brackets
with teh Dirac brackets, we obtain the same equations of motion. So we have
\be
  {\dot q}^\rho = \lbrace q^\rho,H \rbrace_D = \lbrace q^\rho,H \rbrace
  + \lbrace q^\rho, \Phi_{\bar \alpha} \rbrace
  C^{{\bar \alpha}{\bar \beta}} \lbrace H, \Phi_{\bar \beta} \rbrace =
  \lbrace q^\rho , H \rbrace ,
\lbl{3.56}
\ee
where we have taken into account that $\lbrace H, \Phi_{\bar \beta} \rbrace =0$.

To check that we have correctly computed the Dirac brackets 
(\ref{3.44})--(\ref{3.55}), it is instructive to derive
$$
  {\dot q}^\rho = \lb q^\rho,H \rb_D = p_\mu \lb q^\rho,q^\mu \rb_D -
\frac{e^3}{4 \mu} \lb q^\rho,\pi^2 \rb_D - \frac{m}{2e} \lb q^\rho,q^2 \rb_D$$
\be
 \hs{3cm} + \frac{\beta}{e} \lb q^\rho,\pi^\mu \rb_D q_\mu +
   \frac{\beta}{e} \lb q^\rho,q^\mu \rb_D \pi_\mu + \beta \lb q^\rho,p_e \rb_D,
\lbl{3.57}
\ee
and the analogous equations for the other dynamical variables.
Besides Eqs.\,(\ref{3.44})--(\ref{3.46b}), we also need
\be
  \lb q^\rho,p_e \rb_D = \lb q^\rho,\Phi_{\bar \alpha} \rb 
  C^{{\bar \alpha}{\bar \beta}} \lb p_e , \Phi_{\bar \beta} \rb
\lbl{3.58}
\ee
Using (\ref{3.43a}) and
\be
  \lb p_e,\phi_{\bar \alpha} \rb = (0,0,- \frac{3M^2}{e m}+\frac{2m}{e},
  - \frac{M^2}{e}),
\lbl{3.59}
\ee
we have
\be
   \lb q^\rho,p_e \rb_D = \frac{q^\rho}{e} .
\lbl{3.60}
\ee
If we explicitly calculate the terms in Eq.\,(\ref{3.57})
by using (\ref{3.44})--(\ref{3.46b}) and (\ref{3.60}), we obtain
\be
   {\dot q}^\rho = - \frac{e^3}{2 \mu} \pi^\rho + \frac{\beta}{e} q^\rho,
\lbl{3.61}
\ee
which is in agreement with Eq.\,(\ref{3.20}). In deriving the latter
equation we have taken
into account that for the metric $g^{\rho \mu} = \eta^{\rho \mu} -
p^\rho p^\mu/M^2$ we have
\be
  g^{\rho \mu} \pi_\mu = \eta^{\rho \mu} \pi_\mu - p^\rho p^\mu \pi_\mu/M^2
  = \eta^{\rho \mu} \pi_\mu = \pi^\rho,
\lbl{3.61a}
\ee
\be
  g^{\rho \mu} q_\mu = q^\rho - p^\rho \frac{p^\mu q_\mu}{M^2}
  =q^\rho - p^\rho \frac{e}{m} = V^\rho ,
\lbl{3.61b}
\ee
where we have used the constraints $\Phi_3$ and $\Phi_5$, which give
$\pi^\mu p_\mu=0$ and $m p^\mu q_\mu/e = M^2$.

We will now introduce the new dynamical variables,
\be
   \Gam^\mu = \frac{q^\mu}{e}-\frac{p^\mu}{m} = \frac{V^\mu}{e}~,~~~~~~~~~~
   \Pi^\mu = e \pi^\mu ,
\lbl{3.62}
\ee
which are invariant under reparametrizations $\tau \rightarrow \tau' =
h(\tau)$. They satisfy the following relations:
\be
  {\tl S}^{\mu \nu} = \frac{2 \mu (m^2-M^2)}{m} 
  \lbrace \Gam^\mu,\Gam^\nu \rbrace_D
\lbl{3.62a}
\ee
\be
\lbrace \Gam^\rho, {\tl S}^{\mu \nu} \rbrace_D =
   \Gam^\mu g^{\nu \rho} - \Gam^\nu g^{\mu \rho}
\lbl{3.62b}
\ee
\be
  {\tl S}^{\mu \nu} = \frac{(m^2-M^2)}{m^2} 
  \lbrace \Pi^\mu,\Pi^\nu \rbrace_D
\lbl{3.62c}
\ee
\be
\lbrace \Pi^\rho, {\tl S}^{\mu \nu} \rbrace_D =
   \Pi^\mu g^{\nu \rho} - \Pi^\nu g^{\mu \rho}
\lbl{3.62d}
\ee
\be
   \lbrace \Gam^\mu,\Pi^\nu \rbrace = g^{\mu \nu}
   + \frac{m}{m^2-M^2} \left ( \frac{\Pi^\mu \Pi^\nu}{2 \mu} 
   + m \Gam^\mu \Gam^\nu \right ) .
\lbl{3.62e}
\ee
The second class constraints now read as
\bear
  &&\phi_2 = \Pi^\mu (\Gam^\mu + \frac{p^\mu}{m}) \lbl{3.62A}\\
  &&\phi_3 = \Pi^\mu p_\mu \lbl{3.62B}\\
  &&\phi_4 = \frac{\Pi^\mu \Pi_\mu}{2 \mu} - m \Gam^\mu \Gam_\mu -
  p_\mu \Gam^\mu \lbl{3.62C}\\
  &&\phi_5 = - p_\mu \Gam^\mu ,\lbl{3.62D}
\ear
which can be transformed into the following set of constraints:
\bear
  &&\psi_2 = \phi_2 - \frac{\phi_3}{m} = \Pi^\mu \Gam_\mu \lbl{3.62E}\\
  &&\psi_3 = \phi_3 = \Pi^\mu p_\mu \lbl{3.62F}\\
  &&\psi_4 = \phi_4 + \phi_5 = \frac{\Pi^\mu \Pi_\mu}{2 \mu}
   - m \Gam^\mu \Gam_\mu \lbl{3.62G}\\
  &&\psi_5 = - \phi_5 = p_\mu \Gam^\mu. \lbl{3.62H}
\ear 
In terms of the new variables, the Hamiltonian (\ref{3.18}) (for $\alpha=0$)
reads
\be
  H = e \left [-\frac{m}{2} \Gam^\mu \Gam_\mu - \frac{1}{4 \mu} \Pi^\mu \Pi_\mu
  + \frac{1}{2m} (p^2 - m^2) \right ] + \frac{\beta}{e}
   (e p_e +\Pi_\mu \Gam^\mu  )
\lbl{3.63}
\ee
It is a superposition of the first class constraints
\bear
  &&\phi_1 = -\frac{m}{2} \Gam^\mu \Gam_\mu - \frac{1}{4 \mu} \Pi^\mu \Pi_\mu
  + \frac{1}{2m} (p^2 - m^2) \lbl{phi1a}\\
  &&\phi_6 = e p_e +\Pi_\mu \Gam^\mu \lbl{phi6a}
\ear

As an example, let us compute the following quantity, which is invariant under
reparametrizations of $\tau$:
\be
   \frac{{\dot \Gam}^\rho}{e}= \frac{1}{e} \frac{\dd}{\dd \tau}
   \left (\frac{q^\rho}{e} \right )
    = \frac{1}{e} \lb \Gam^\rho,H \rb_D
\lbl{3.63a}
\ee
Because $\lb \Gam^\rho,p^2 \rb_D =0$ and $\lb \Gam^\rho, 
\frac{\beta}{e} (e p_e +\Pi_\mu \Gam^\mu  ) \rb_D =0$, we have
\bear
   \frac{{\dot \Gam}^\rho}{e}&=& \lb \Gam^\rho, - \frac{m}{2} \Gam^\mu \Gam_\mu
   - \frac{1}{4 \mu} \Pi^\mu \Pi_\mu \rb_D =
   -\frac{m}{2} \lb \Gam^\rho,\Gam^\mu \Gam_\mu \rb_D 
   - \frac{1}{4 \mu} \lb \Pi^\mu \Pi_\mu \rb_D \nonumber \\
  &=& -\frac{m}{2} \left [ \Gam_\mu \lb \Gam^\rho,\Gam^\mu \rb_D
   +\lb \Gam^\rho,\Gam^\mu \rb_D \Gam_\mu \right ]
   - \frac{1}{4 \mu} \left [ \Pi_\mu \lb \Gam^\rho,\Pi^\mu \rb_D
   + \lb \Gam^\rho,\Pi^\mu \rb_D \Pi_\mu \right ] \nonumber \\
   &=& -\frac{m}{2} \frac{m}{2 \mu (m^2-M^2)}(\Gam_\mu {\tl S}^{\rho \mu}
   +{\tl S}^{\rho \mu} \Gam_\mu )\nonumber \\   
   &&\hs{4cm}- \frac{2}{4 \mu} \Pi_\mu \left [ g^{\rho \mu}+ \frac{1}{2(m^2-M^2)}
   \left ( \frac{m}{\mu} \Pi^\rho \Pi^\mu + 2 m^2 \Gam^\rho \Gam^\mu \right )
   \right ] \nonumber \\
   &=& -\frac{m}{2} \frac{2 m}{2 \mu (m^2-M^2)} \Gam_\mu (\Gam^\rho \Pi^\mu -
   \Gam^\mu \Pi^\rho) - \frac{1}{2 \mu} \left [ \Pi_\mu g^{\rho \mu}
   + \frac{1}{2(m^2-M^2)}\frac{m}{\mu} \Pi^\rho \Pi_\mu \Pi^\mu \right ]
   \nonumber \\
   &=& -\frac{1}{2 \mu} \Pi^\rho + \frac{m \Pi^\rho}{2 \mu (m^2-M^2)}
   +\left (  m \Gam^\mu \Gam_\mu - \frac{\Pi^\mu \Pi_\mu}{2 \mu}  \right ) .
\lbl{3.64}
\ear
In the last term of the latter equation we have the expression
\be
  m \Gam^\mu \Gam_\mu - \frac{\Pi^\mu \Pi_\mu}{2 \mu} 
  = m \left (\frac{q^\mu}{e}-\frac{p^\mu}{m} \right ) 
    \left (\frac{q_\mu}{e}-\frac{p_\mu}{m} \right ) - \frac{e^2 \pi^2}{2 \mu}
    =0,
\lbl{3.65}
\ee
which vanishes because of the constraints $\Phi_4=0$ and $\Phi_5=0$.
Therefore, Eq.\,(\ref{3.64}) gives
\be
\frac{{\dot \Gam}^\rho}{e}= \lb \Gam^\rho, - \frac{m}{2} \Gam^\mu \Gam_\mu
   - \frac{1}{4 \mu} \Pi^\mu \Pi_\mu \rb_D = -\frac{1}{2 \mu} \Pi^\rho .
\lbl{3.66}
\ee

The latter result can be much quicker obtained if instead of the
Dirac brackets we use the Poisson brackets, which satisfy
\be
     \lb \Gam^\rho, \Gam_\mu \rb = 0, ~~~~~~~~~~\lb \Gam^\rho, \Pi^\mu \rb
     = \eta^{\rho \mu}.
\lbl{3.67}
\ee
Then it is straightforward to verify that
\be
\frac{{\dot \Gam}^\rho}{e}= \lb \Gam^\rho, - \frac{m}{2} \Gam^\mu \Gam_\mu
   - \frac{1}{4 \mu} \Pi^\mu \Pi_\mu \rb = -\frac{1}{2 \mu} \Pi^\rho .
\lbl{3.68}
\ee

Similarly, we obtain
\be
   \frac{{\dot \Pi}^\rho}{e} = m \Gam^\rho .
\lbl{3.68a}
\ee
Together, Eqs.\,(\ref{3.68}),(\ref{3.68a}) give
\be
   \frac{\dd^2 \Gam^\rho}{\dd s^2} + \omega^2 \Gam^\rho = 0,
\lbl{3.68b}
\ee
where $\omega^2 = m^2/(2 \mu)$, and $\dd s= e \dd \tau$.

However, the longer procedure with the Dirac brackets provides a test for
the correctness of the computed Dirac brackets. It will also
guide us in the quantized theory in which the Dirac
brackets will be replaced by commutators.

\section{Quantization}

We will now consider the quantization based on the Dirac brackets. 
According to this procedure, the classical quantities become the operators that
satisfy the commutation relations corresponding to the Dirac bracket
relations\,\ci{Dirac}. In the following, we will consider the quantum versions
of the Dirac bracket relations of the previous section.

{\it The first class constraints} $\phi_1$ and $\phi_6$ act as restrictions
on the Hilbert space of states according to
\be
   \phi_1 |\psi \rangle = 0~,~~~~~~~~~~~~~~~\phi_6 |\psi \rangle = 0,
\lbl{4.1}
\ee
where the states satisfying the above equations are {\it physical states}.

{\it The second class constraints} commute with all phase space operators and any
function of them,
\be
   [\phi_{\bar \alpha}, f(x^\mu,p_\mu,q^\mu,\pi_\mu,e,p_e)] = 0.
\lbl{4.2}
\ee
Therefore, they are $c$-numbers that satisfy
\be
    \phi_{\bar \alpha} = 0~,~~~~~~~~~~{\bar \alpha} = 2,3,4,5 .
\lbl{4.3}
\ee

Let us introduce the operators 
\be
\Gam^\mu = \frac{V^\mu}{e}, ~~~~~~~~~ {\Pi}^\mu = e \pi^\mu,
\lbl{4.6}
\ee
whose classical analogs, satisfying the constraints
(\ref{3.62E})--(\ref{3.62H}),
were introduced in Eq.\,(\ref{3.62}). The corresponding quantum constraints
have the same form, except that $\psi_2$ is now symmetrized according to
\be
  \psi_2 = \frac{1}{2} (\Pi^\mu \Gam_\mu + \Gam_\mu \Pi^\mu)
\lbl{4.6a}
\ee

We will now consider the quantum version of
the Dirac bracket relations (\ref{3.62a})--(\ref{3.62e}),
satisfied by $\Gam^\mu$ and $\Pi^\mu$. Writing
\be
   \frac{2 \mu (m^2-p^2)}{m} \equiv \frac{\rho^2}{4},
   ~~~~~~~~~\frac{(m^2-p^2)}{m^2} \equiv \frac{{\tl \rho}^2}{4}
\lbl{4.4}
\ee
\be
  g^{\mu \nu} = \eta^{\mu \nu} - \frac{1}{p^2} p^\mu p^\nu
\lbl{4.5}
\ee
we have
\be
  {\tl S}^{\mu \nu} = j_1 \frac{\rho^2}{4} [\Gam^\mu,\Gam^\nu ],
\lbl{4.7}
\ee
\be
  [\Gam^\rho, {\tl S}^{\mu \nu} ] 
  = j_2 (\Gam^\mu g^{\nu \rho} - \Gam^\nu g^{\mu \rho}).
\lbl{4.8}
\ee
\be
  {\tl S}^{\mu \nu} = j_3 \frac{{\tl \rho}^2}{4} 
  [{\Pi}^\mu,{\Pi}^\nu ],
\lbl{4.9}
\ee
\be
  [{\Pi}^\rho, {\tl S}^{\mu \nu} ] 
  = j_4 ({\Pi}^\mu g^{\nu \rho} - {\Pi}^\nu g^{\mu \rho}).
\lbl{4.10}
\ee
The quantum version of the Dirac bracket (\ref{3.48}) is
\be
   [\Gam^\mu,{\Pi}^\nu] = j \left ( g^{\mu \nu} + \frac{4}{\rho^2}
   {\Pi}^\mu \cdot {\Pi}^\nu + \frac{4}{{\tl \rho}^2}
   \Gam^\mu \cdot \Gam^\nu \right ),
\lbl{4.10a}
\ee
where the dot denotes the symmetrized product, e.g.,
$\Gam^\mu \cdot \Gam^\nu = \frac{1}{2} (\Gam^\mu \Gam^\nu 
+ \Gam^\nu \Gam^\mu )$.

In addition we also have the relation
\be
  {\tl S}^{\mu \nu} = \Gam^\mu {\Pi}^\nu - \Gam^\nu {\Pi}^\mu
\lbl{4.10b}
\ee
which corresponds to the definition ${\tl S}^{\mu \nu}
 = V^\mu \pi^\nu - V^\nu \pi^\mu$ of the modified spin tensor
 ${\tl S}^{\mu \nu}$.

The relations (\ref{4.7})--(\ref{4.10b}) are the quantum counterpart of
the classical equations (\ref{3.62a})--(\ref{3.62e}),(\ref{3.46}), whose
structure is more complicated than that of the usual
relations, such as $\lbrace x^\mu,p_\nu \rbrace = {\delta^\mu}_\nu$,
$\lbrace x^\mu,x^\nu \rbrace = 0$, $\lbrace p_\mu,p_\nu \rbrace = 0$.
The latter Poisson bracket relations can be replaced by the operator
commutation relations $[{\hat x}^\mu, {\hat p}_\nu ] = {\delta^\mu}_\nu$,
$[{\hat x}^\mu,{\hat x}^\nu ]=0$, $[{\hat p}_\mu,{\hat p}_\nu]=0$, that
are satisfied by the operators represented as ${\hat x}^\mu = x^\mu$,
${\hat p}_\mu = - \p_\mu$. However, the operator ${\hat p}_\mu$ so
defined, is not Hermitian, therefore we make the replacement
${\hat p}_\mu = -\p_\mu \rightarrow {\hat p}_\mu = - i \p_\mu$, so that the
quantum commutator becomes $[{\hat x}^\mu, {\hat p}_\nu ] = i {\delta^\mu}_\nu$.
In  the case of the rather complicated Poisson bracket system
(\ref{3.46a}),(\ref{3.46b}),(\ref{3.53}),(\ref{3.55}),(\ref{3.50}),
or equivalently, (\ref{3.62a})--(\ref{3.62e}),
we cannot a priori
expect that the replacement $\lbrace ~,~ \rbrace_D \rightarrow
\frac{1}{i} [~,~]$ will work. Therefore, in the commutation relations
(\ref{4.7})--(\ref{4.10a}) we have introduced the quantities
$j_1$,$j_2$,$j_3$,$j_4$ and $j$, which will be determined in the process
of finding a consistent representation for $\Gam^\mu$, ${\Pi}^\mu$,
satisfying  those commutation relations, as well as the definition (\ref{4.10b})
of ${\tl S}^{\mu \nu}$.

The solution to the system (\ref{4.7})--(\ref{4.10}) are the
operators $\Gam^\mu$, ${\Pi}^\mu$ that satisfy the Clifford
algebra relations with the modified metric:
\be
   \frac{j_1}{j_2}\frac{1}{2} (\Gam^\mu \Gam^\nu + \Gam^\nu \Gam^\mu)
    =  - \rho^{-2} g^{\mu \nu} ,
\lbl{4.11}
\ee
\be
   \frac{j_3}{j_4}\frac{1}{2} ({\Pi}^\mu {\Pi}^\nu 
   + {\Pi}^\nu {\Pi}^\mu) 
    =  - {\tl \rho}^{-2} {g}^{\mu \nu} .
\lbl{4.12}
\ee

The metric $g^{\mu \nu}=\eta^{\mu \nu} - p^\mu p^\nu/p^2$ can be transformed from the Minkowski metric
$\eta^{\mu \nu}$ $= {\rm diag} (1,-1,-1,-1)$ according to
\be
  {g}^{\mu \nu} = \frac{\p x'^\mu}{\p x^\alpha}\frac{\p x'^\nu}{\p x^\beta} 
   \eta^{\alpha \beta},
\lbl{4.13}
\ee
where 
\be
    x' =  \left (x^\mu - \frac{p^\mu x^\rho p_\rho}{p^2} \right )
    ~,~~~~~~ \frac{\p x'^\mu}{\p x^\alpha} 
    = \left ( {\delta^\mu}_\alpha 
    -  \frac{p^\mu p_\alpha}{p^2} \right ) .
\lbl{4.14}
\ee
We see that because of the factors $\rho^2$ and ${\tl \rho}^2$
in Eqs.\,(\ref{4.11}) and (\ref{4.12}), the above
coordinate transformation is accompanied by the corresponding dilatation.

Eqs.\,(\ref{4.7}),(\ref{4.8}) (and (\ref{4.9}),(\ref{4.10})) are similar
to the equations that come from the relativistic covariance of
the Dirac equation, which gives
\be
  [\gam^\rho, \sigma^{\mu \nu}] = 2 i (\eta^{\rho \mu} \gam^\nu -
  \eta^{\rho \nu} \gam^\mu],
\lbl{4.14a}
\ee
where $\sigma^{\mu \nu}$ are generators of Lorentz transformations.
The latter equation is satisfied by
\be
  \sigma^{\mu \nu} = \frac{i}{2} [\gam^\mu,\gam^\nu].
\lbl{4.14b}
\ee
In the theory of the Dirac equation, the Clifford algebra relations for
the objects $\gam^\mu$, namely, 
\be
\gam^\mu \gam^\nu + \gam^\nu \gam^\mu
= 2 \eta^{\mu \nu},
\lbl{4.14c}
\ee
are already given, whereas the relation (\ref{4.14b}) for the generators
$\sigma^{\mu \nu}$ is computed from (\ref{4.14a}) and (\ref{4.14c}).
In our procedure, on the contrary, we arrived at the commutation relations
(\ref{4.7}), (\ref{4.8}), which are very similar to
Eqs.\,(\ref{4.14b}),(\ref{4.14a}), respectively,
and a question was, how to represent the operators $\Gam^\mu$. We have found by
direct calculation that $\Gam^\mu$ satisfy the Clifford algebra relations 
(\ref{4.11}). Analogous hold for the operators $\Pi^\mu$, satisfying
Eqs.\,(\ref{4.9}),(\ref{4.10}),(\ref{4.12}).

\subsection{Schr\"odinger picture}

In the Schr\"odinger picture, the operators do not evolve in time.

If $j_1=j_2$, then Eq.\,(\ref{4.11}) is satisfied by
\be
   \Gam^\mu = \frac{e_q}{\rho}\frac{\p x'^\mu}{\p x^\alpha} \gam^\alpha =
   \frac{e_q}{\rho} \left ( \gam^\mu - \frac{p^\mu p_\alpha \gam^\alpha}{p^2}
   \right ),
\lbl{4.15}
\ee
where $e_q^2 =-1$, whereas $\gam^\mu$ satisfy the Clifford algebra
relations (\ref{4.14c}).

Similarly, if $j_3 = j_4$, then Eq.\,(\ref{4.12}) is satisfied by\footnote{
We take the minus sign in front of $e_\pi$ for later convenience..}
\be
   \Pi^\mu = -\frac{e_\pi}{\tl \rho}\frac{\p x'^\mu}{\p x^\alpha} \gam^\alpha =
   -\frac{e_\pi}{\tl \rho} \left ( \gam^\mu - \frac{p^\mu p_\alpha \gam^\alpha}{p^2}
   \right ),
\lbl{4.17}
\ee
where $e_\pi^2 = -1$. We assume that $e_q$ and $e_\pi$ commute with $\gam^\mu$,

The quantities
\be
   \alpha^\mu = \gam^\mu - \frac{p^\mu p_\alpha \gam^\alpha}{p^2}
\lbl{4.18}
\ee
are projections of $\gam^\mu$ onto the 3-dimensional hypersurface that is
orthogonal to the direction of the 4-momentum $p^\mu$. Thus, vectors
$\alpha^\mu$ are ``spatial" parts of vectors $\gam^\mu$. They satisfy the
Clifford algebra relations
\be
   \alpha^\mu \cdot \alpha^\nu \equiv \frac{1}{2} (\alpha^\mu \alpha^\nu
   + \alpha^\nu \alpha^\mu) = g^{\mu \nu}
    = \eta^{\mu \nu} - \frac{p^\mu p^\nu}{p^2}
\lbl{4.19}
\ee

In view of the above relations (\ref{4.15})--(\ref{4.18}), the spin operator
determined by Eq.\,(\ref{4.7}) and (\ref{4.9}), becomes
\be
    {\tl S}^{\mu \nu} = - \frac{j_1}{4} [\alpha^\mu, \alpha^\nu]
    = - \frac{j_3}{4} [\alpha^\mu, \alpha^\nu] ,
\lbl{4.20}
\ee
from which it follows that $j_1 = j_3$.

On the other hand, according to Eq.\,(\ref{4.10b}), the spin tensor is
\be
   {\tl S}^{\mu \nu} = \Gam^\mu \Pi^\nu - \Gam^\nu \Pi^\mu =
   -\frac{e_q e_\pi}{\rho {\tl \rho}} [\alpha^\mu,\alpha^\nu] .
\lbl{4.21}
\ee
By comparing Eqs.\,(\ref{4.20}) and (\ref{4.21}) we obtain the relations
\be
  j_1 = e_q e_\pi
\lbl{4.22}
\ee
\be
    \rho {\tl \rho} = 4~,~~~~~~{\rm i.e.},~~~~m^2-p^2 
    - m \sqrt{\frac{m}{2 \mu}}=0 ,
\lbl{4.23}
\ee
where $p^2 \equiv p^\mu p_\mu$ is the squared momentum operator.

Using (\ref{4.15}),(\ref{4.17}), we have
\be
   [\Gam^\mu,\Pi^\nu] = -\frac{1}{\rho {\tl \rho}} 
   [e_q \alpha^\mu,e_\pi \alpha^\nu ] = -\frac{1}{\rho {\tl \rho}}
    (e_q e_\pi \alpha^\mu \alpha^\nu - e_\pi e_q \alpha^\nu \alpha^\mu).
\lbl{4.24}
\ee
On the other hand, according to (\ref{4.10a}), the commutator is
\be
  [\Gam^\mu,\Pi^\nu] = j \left ( g^{\mu \nu} + \frac{4}{\rho^2}
   {\Pi}^\mu \cdot {\Pi}^\nu + \frac{4}{{\tl \rho}^2}
   \Gam^\mu \cdot \Gam^\nu \right ) = j g^{\mu \nu}
   \left ( 1 - \frac{8}{\rho^2 {\tl \rho}^2} \right ) = j \frac{g^{\mu \nu}}{2} ,
\lbl{4.25}
\ee
where in the last step we used Eq.\,(\ref{4.23}).

The commutators in Eqs.\,(\ref{4.24}) and (\ref{4.25}) must
be the same. This is possible, if $e_q$ and $e_\pi$ are the numbers
satisfying, besides
\be
e_q^2=-1~,~~~~~~~ e_\pi^2=-1,
\lbl{4.25a}
\ee
 also the relation
\be
   e_q e_\pi + e_\pi e_q =0.
\lbl{4.26}
\ee
This means that $e_q$, $e_\pi$ are elements of the Clifford algebra
$Cl(0,2)$. The latter notation means that the signature of the vector
space spanned by the basis vectors $e_q$ and $e_\pi$
is $(- -)$ or $(0,2)$.\footnote{In general, $Cl(p,q)$ is the Clifford
algebra of a vector space with signature $(p,q)$. In particular, the
Clifford algebra of spacetime is $Cl(1,3)$.}

Taking into account Eq.\,(\ref{4.26}) in Eq.\,(\ref{4.24}), we obtain
\be
  [\Gam^\mu,\Pi^\nu] = -\frac{1}{\rho {\tl \rho}}
  e_q e_\pi (\alpha^\mu \alpha^\nu +\alpha^\nu \alpha^\mu) =
 - e_q e_\pi \frac{2 g^{\mu \nu}}{\rho {\tl \rho}} .
\lbl{4.27}
\ee
By using in the latter equation the relation (\ref{4.23}), we arrive
at the same result as in Eq.\,(\ref{4.25}), if we put
\be
    j= -e_q e_\pi .
\lbl{4.28}
\ee
Altogether, we have
\be
   j_1=j_2=j_3=j_4=-j = e_q e_\pi.
\lbl{4.28a}
\ee
The square is $j^2 =(e_q e_\pi)^2 = e_q e_\pi e_q e_\pi = - e_q e_\pi^2 e_q =
e_q^2 = -1$. We see that $j$, which is a bivector of the Clifford algebra
$Cl(0,2)$, behaves as  imaginary unit. 

Thus we have found that the system of the operator equations
(\ref{4.7})--(\ref{4.10b}) is solved by the operators $\Gam^\mu$,
$\Pi^\mu$, defined according to Eqs.\,(\ref{4.15})--(\ref{4.17}),
provided that the quantities $j_1$,$j_2$,$j_3$,$j_4$ and $j$ satisfy
(\ref{4.28a}).

From (\ref{4.20})--(\ref{4.18}) we have the following expression for the
spin operators
\be
   {\tl S}^{\mu \nu} = - \frac{j_1}{4}[\gam^\alpha,\gam^\beta]
   {g_\alpha}^\mu {g_\beta}^\nu ,
\lbl{4.29}
\ee
where ${g_\alpha}^\mu = {\delta_\alpha}^\mu - \frac{p_\alpha p^\mu}{p^2}$.
The Pauli-Lubanski operator in 4-dimensions is
\be
  S_\mu = \mbox{$\frac{1}{2 M}$} \epsilon_{\mu \nu \rho \sigma}
  {\tl S}^{\nu \rho} p^\sigma 
  = \mbox{$\frac{1}{2 M}$}\epsilon_{\mu \nu \rho \sigma} S^{\nu \rho}p^\sigma,
\lbl{4.30}
\ee
where
\be
   S^{\nu \rho} = - \frac{j_1}{4} [\gam^\nu, \gam^\nu]
\lbl{4.31}
\ee
is the usual spin operator, given in terms of the Dirac gammas. The eigenvalues
of its square, $S^\mu S_\mu$, are $s(s+1)= 3/4$, implying $s=\frac{1}{2}$.

\subsection{Heisenberg picture}

In the quantized theory the classical Hamiltonian is replaced by the
corresponding operator. Instead of (\ref{3.63a}), we have the
Heisenberg equations of motion
\be
   \frac{{\dot \Gam}^\rho}{e} = - j \frac{1}{e} [\Gam^\rho,H]
   = - j [\Gam^\rho, - \frac{m}{2} \Gam^\mu \Gam_\mu
   - \frac{1}{4 \mu} \Pi^\mu \Pi_\mu].
\lbl{4.32}
\ee
In the following we will use the relations
\be
   [\Gam^\rho,\Gam^\mu \Gam_\mu] = [\Gam^\rho,\Gam^\mu] \Gam_\mu
   +\Gam_\mu [\Gam^\rho,\Gam^\mu] ,
\lbl{4.33}
\ee
\be
   [\Gam^\rho, \Pi^\mu \Pi_\mu] = [\Gam^\rho,\Pi^\mu] \Pi_\mu
   + \Pi_\mu [\Gam^\rho,\Pi^\mu].
\lbl{4.34}
\ee
Inserting
\be
  [\Gam^\rho,\Gam^\mu] = - \frac{4}{j \rho^2} {\tl S}^{\rho \nu} ~,
  ~~~~~~~j_1 = - j = e_q e_\pi,
\lbl{4.35}
\ee
and
\be
    [\Gam^\rho,\Pi^\mu] = j \left ( g^{\rho \mu} + \frac{4}{\rho^2}
    \Pi^\rho \cdot \Pi^\mu + \frac{4}{{\tl \rho}^2} \Gam^\rho \cdot \Gam^\mu
    \right ) \equiv j {\tl G}^{\rho \mu}
\lbl{4.36}
\ee
into (\ref{4.32}), we obtain
\be
    \frac{{\dot \Gam}^\rho}{e} = - \frac{m}{2} (\Gam_\mu {\tl S}^{\rho \mu}
    +{\tl S}^{\rho \mu} \Gam_\mu ) \frac{4}{j \rho^2} +
   \frac{1}{4 \mu} (\Pi_\mu {\tl G}^{\rho \mu} +{\tl G}^{\rho \mu} \Pi_\mu ),
\lbl{4.37}
\ee
which, after using the definition (\ref{4.10b}) of ${\tl S}^{\rho \mu}$
and the constraint (\ref{4.6a}), becomes
\be
    \frac{{\dot \Gam}^\rho}{e} = - \frac{1}{2 \mu} \left ( 2 \Pi_\mu 
     g^{\rho \mu} + \frac{4}{\rho^2} \Pi_\mu \Pi^\rho \Pi^\mu \right )
     - \frac{2 m}{\rho^2} \left (\Gam_\mu \Gam^\rho \Pi^\mu - \Gam_\mu \Pi^\rho
     \Gam^\mu + \Pi^\mu \Gam^\rho \Gam_\mu \right ).
\lbl{4.38}
\ee
The latter equation, apart form the order of operators, matches the
classical equation (\ref{3.64}). By using (\ref{4.11}),(\ref{4.12}) and
(\ref{4.36}), we can reverse the order of operators in the
products $\Pi^\rho \Pi^\mu$, $\Gam^\rho \Pi^\mu$, $\Pi^\rho \Gam^\mu$ and
$\Gam^\rho \Gam_\mu$, at the expense of acquiring certain extra terms.
Using also the operator version of the constraint (\ref{3.62G}), we arrive
at the equation
\be
    \frac{{\dot \Gam}^\rho}{e} = - \frac{1}{2 \mu} \Pi^\rho +
    \frac{4}{\mu \rho^2 {\tl \rho}^2} \Pi^\rho 
    - j \frac{4 m}{\rho^2} {\tl G}^{\rho \mu} \Gam_\mu
\lbl{4.39}
\ee
The latter equation, in comparison with the corresponding classical
equation (\ref{3.66}), has two extra terms.

Similarly, we obtain
\be
   \frac{{\dot \Pi}^\rho}{e} 
   = m \Gam^\rho - \frac{8m}{\rho^2 {\tl \rho}^2} \Gam^\rho 
   + j \frac{2}{\mu {\tl \rho}^2} {\tl G}^{\rho \mu} \Pi_\mu .
\lbl{4.40}
\ee

In the absence of the extra terms, Eqs.,(\ref{4.39}),(\ref{4.40}) would
give
\be
   \frac{\dd^2 \Gam^\rho}{\dd s^2} + \omega^2 \Gam^\rho = 0~,~~~~~~
   \Pi^\rho = - 2 \mu \frac{\dd \Gam^\rho}{\dd s}~ , ~~~~~~
   \dd s = e \dd \tau,
\lbl{4.41}
\ee
with the solution
\be
   \Gam^\rho = a^\rho {\rm cos} \, \omega s + b^\rho {\rm sin} \, \omega s,
\lbl{4.42}
\ee
\be
   \Pi^\rho = - 2 \mu \omega (- a^\rho {\rm sin} \, \omega s +
   b^\rho {\rm cos} \, \omega s ) ,
\lbl{4.43}
\ee
where
\be
   \omega = \sqrt{\frac{m}{2 \mu}}.
\lbl{4.44}
\ee

At $s=0$, we have
\be
  \Gam^\rho (0) = a^\rho~,~~~~~~~\Pi^\rho (0) = - 2 \mu \omega b^\rho
\lbl{4.45}
\ee
Comparison with Eqs.\,(\ref{4.15}),(\ref{4.17}) gives
\be
   a^\rho = \frac{e_q \alpha^\rho}{\rho}~,~~~~~~~~~
   b^\rho = \frac{e_\pi \alpha^\rho}{\rho},
\lbl{4.46}
\ee
where $\alpha^\rho$ is defined in Eq.\,(\ref{4.18}).

If we insert the expressions (\ref{4.46}) for $a^\rho$ and $b^\rho$
into Eqs.\,(\ref{4.42}),(\ref{4.43}), we obtain the following relations
between $\Pi^\rho$ and $\Gam^\rho$:
\be
   \Pi^\rho = - 2 \mu \omega e_q e_\pi \Gam^\rho .
\lbl{4.47}
\ee
We see that the bivector $e_q e_\pi$ performs a $\pi/2$ rotation in
phase space, and thus, up to the factor $2 \mu \omega$, exchanges
$\Pi^\rho$ and $\Gam^\rho$.

 In deriving Eq.\,(\ref{4.47}), we assumed that the
second and the third term in the equations of motion (\ref{4.39}),
(\ref{4.40}) cancel out. According to Eq.\,(\ref{4.39}) this is the case if
\be
  \Pi^\rho = j \mu m {\tl \rho}^2 \, \Gam_\mu {\tl G}^{\rho \mu} ,
\lbl{4.48}
\ee
and according to eq.\,(\ref{4.40}) if
\be
  \Gam^\rho= - \frac{j \rho^2}{4 \mu m} {\tl G}^{\rho \mu} \Pi_\mu .
\lbl{4.48a}
\ee
Using Eqs.\,(\ref{4.36}),(\ref{4.23}),(\ref{4.11}),(\ref{4.12}), we obtain
\be
   {\tl G}^{\rho \mu} = \frac{g^{\rho \mu}}{2},
\lbl{4.49}
\ee
and
\be
  \Pi^\rho = j \sqrt{2 m \mu}\, \Gam^\rho,
\lbl{4.50}
\ee
which gives the relations (\ref{4.47}), if we insert $j=- e_q e_\pi$ and
$\omega = \sqrt{m/2\mu}$. The cancelation of the second and the third
term in Eqs.\,(\ref{4.39}),(\ref{4.40}) is thus consistent with the
equations of motion (\ref{4.41}) that give the relation (\ref{4.47}).

The Heisenberg equations of motion (\ref{4.37}) are thus consistent with
the representation of operators (\ref{4.15}),(\ref{4.17}), evolving
according to Eqs.\,(\ref{4.42},(\ref{4.43}).

\subsection{The physical states}

The explicit form of Eq.\,(\ref{4.1}) is the quantum analog of the constraints
$\phi_1$,$\phi_6$ given in Eqs.\,(\ref{3.30}),(\ref{3.26}), or equivalently,
in Eqs.\,(\ref{phi1a}),(\ref{phi6a}). The physical states thus satisfy  the
equation
\be
  \left [ - \frac{m}{2} \Gam^\mu \Gam_\mu - \frac{1}{4 \mu} \Pi^\mu \Pi_\mu
  + \frac{1}{2 m} (p^2 - m^2) \right ] |\Psi \rangle = 0,
\lbl{4.51}
\ee
\be
  (e p_e + \Pi_\mu \Gam^\mu ) |\Psi \rangle = 0.
\lbl{4.52}
\ee

Because of the second class constraint $\psi_2 = \Pi_\mu \Gam^\mu = 0$,
Eq.\,(\ref{4.52}) gives $p_e |\Psi \rangle = 0$, which is in agreement
with the classical equation $p_e =0$.

If we use the Clifford algebra relations (\ref{4.11}),(\ref{4.12}) and
Eqs.\,(\ref{4.4}),(\ref{4.5}), we obtain
\be
  m \Gam^\mu \Gam_\mu = - \frac{(D-1) m^2}{8 \mu (m^2-p^2)}
  = \frac{\Pi^\mu \Pi_\mu}{2 \mu} ,
\lbl{4.53}
\ee
where $D$ is the dimension of spacetime. On the other hand, we also
have the relation (\ref{4.23}) which must now be written as a condition
on physical states:
\be
  \left ( m^2-p^2 - m \sqrt{\frac{m}{2 \mu}} \right ) |\psi \rangle .
\lbl{4.54}
\ee
A physical state is a superposition of the basis states, $|x,\alpha \rangle =
|x \rangle |\alpha \rangle$, which can be written as the product of
the coordinate states $|x \rangle$ and the spinor states $|\alpha \rangle$.
With respect to the spin states $|\alpha \rangle$, the operators $p^2$
is diagonal, which justifies usage of the operator equation (\ref{4.23}).
But now we also take into account the coordinate states $|x \rangle$,
in which the operator $p^2$ is not diagonal. Therefore, we have to use
equation (\ref{4.54}). Usage of Eq.\,(\ref{4.23}) was just a short cut,
valid for the matrix elements of $p^2$ between the 
spinor state $|\alpha \rangle$, and so we represented $p^2 \rightarrow
\langle \alpha |p^2 |\alpha' \rangle = p^2 \delta_{\alpha \alpha'}$.

If we plug Eqs.\,(\ref{4.53}) into the equation of state
(\ref{4.51}), and take into account (\ref{4.54}), we find that the terms
do not cancel out, but that there remain the residual terms, giving
\be
  \frac{1}{4} (D-3) \omega|\Psi \rangle = 0 ,
\lbl{4.55}
\ee
where
\be
\omega = \sqrt{\frac{m}{2 \mu}} .
\lbl{4.56}
\ee
The above equation is satisfied if the dimension of spacetime is $D=3$.
Such a restriction on dimensionality of spacetime is very undesirable, and
ruins our construction in which we assumed at least 4-dimensional
spacetime. Alternatively, Eq.\,(\ref{4.55}) is satisfied if $\omega$, which
corresponds to the classical frequency of circular motion, vanishes.
Such a restriction, of course, also invalidates our model.

A possible solution to such an anomaly in the quantized theory is in
introducing yet another time-like dimension into our model.
Let us  generalize the string action to include a $(D+2)^{\rm th}$
dimension which, as the  $(D+1)^{\rm th}$ one, is time-like:
\be
   I = \frac{T}{2} \int \dd \tau \dd \sigma \sqrt{\gam} \gam^{ab}
   \p_a X^{{\hat{\hat \mu}}} \p_b X_{{\hat {\hat \mu}}} .
\lbl{4.57}
\ee
The embedding functions can be split according to
\be
  X^{\hat {\hat \mu}} = (X^{\hat \mu}, X^{D+2})~,
  ~~~~{\hat \mu} = 0,1,2,...,{D-1},{D+1}~,
\lbl{4.58}
\ee
where $X^{\hat \mu}$ occur in the action (\ref{2.9a}), whereas
$X^{D+2}$ is due to the additional dimension. Then we have
\be
  \p_a X^{{\hat{\hat \mu}}} \p_b X_{{\hat {\hat \mu}}}
  = \p_a X^{\hat \mu} \p_b X_{\hat \mu}+\p_a X^{D+2} \p_b X_{D+2} .
\lbl{4.59}
\ee

Inserting (\ref{4.59}) into the action (\ref{4.57}), we obtain
\be
 I = \frac{T}{2} \int \dd \tau \dd \sigma \sqrt{\gam} \gam^{ab}
  (\p_a X^{\hat \mu} \p_b X_{\hat \mu}+\p_a X^{D+2} \p_b X_{D+2}).
\lbl{4.60}
\ee
Variation of the action (\ref{4.60}) with respect to $\gam_{ab}$ gives
\be
  \gam_{ab} = f_{ab} \equiv \p_a X^{{\hat{\hat \mu}}} \p_b X_{{\hat {\hat \mu}}}
  = \p_a X^{\hat \mu} \p_b X_{\hat \mu}+\p_a X^{D+2} \p_b X_{D+2} .
\lbl{4.61}
\ee

Let us assume the following dependence\footnote{Because of the reparametrization
invariance, we are free to choose two of the $D+2$ functions
$X^{{\hat{\hat \mu}}} (\tau,\sigma)$. We have already chosen $X^{D+1}= \sigma$.
Now we we make a choice for $X^{D+2}$.}
 of the extra variable $X^{D+2}$ on $\xi^a=(\tau, \sigma)$:
\be
  X^{D+2} (\tau,\sigma) = K^2 \tau ,
\lbl{4.61a}
\ee
so that
\be
  \p_a X^{{\hat{\hat \mu}}} \p_b X_{{\hat {\hat \mu}}} =
  \p_a X^{\hat \mu} \p_b X_{\hat \mu} + K^2 .
\lbl{4.62}
\ee
Then we have
\be
   f_{11} =\p_1 X^{\hat \mu} \p_1 X_{\hat \mu}
    + K^2 ~, 
    ~~~~f_{12}=\p_1 X^{\hat \mu} \p_2 X_{\hat \mu}~,
    ~~~~f_{22}=\p_2 X^{\hat \mu} \p_2 X_{\hat \mu}
\lbl{4.64}
\ee
which means that only $f_{11}$ is modified by the presence of the
additional, $(D+2)^{\rm th}$, dimension. Instead of Eq.\,(\ref{2.6a}),we have
\be
     f_{ab}=    \begin{pmatrix} {\dot x}^2 +K^2 + 2 {\dot x}{\dot y} k \sigma,
     &   {\dot x} y k\\
                 {\dot x} y k ,      &   k^2 y^2 + \epsilon \\
      \end{pmatrix} ,
\lbl{4.65a}
\ee
In the action (\ref{2.10}), ${\dot x}^2$ should be replaced by
${\dot x}^2 +K^2$. Such a modified action (\ref{2.10}) gives the
same equations of motion (\ref{2.17}),(\ref{2.18}),(\ref{2.19}).
Only Eq.\,(\ref{2.16}), which contains ${\dot x}^2$, is modified.
In view of (\ref{4.64}), the extra term in the action (\ref{4.60}) is
then
\be
  I_1 = \frac{T}{2} \int \dd \tau \dd \sigma \sqrt{\gam} \gam^{11} K^2
  = \frac{LT}{2} \int \dd \tau \frac{K^2}{e} + {\cal O}(k^2 L^2) .
\lbl{4.65}
\ee
It contributes only to the $e$ equation of motion. By repeating the
calculations of Sec.\,3 with inclusion of the extra term (\ref{4.65}),
we arrive at the following modified 1$^{\rm st}$ class constraint:
\be
  \phi_1 = - \frac{m}{2} \Gam^\mu \Gam_\mu - \frac{\Pi^\mu \Pi_\mu}{4 \mu}
  + \frac{1}{2m} (p^2-m^2)-\frac{Q}{2}  ,
\lbl{4.66}
\ee
where
\be
   Q = \frac{m K^2}{e^2} .
\lbl{4.66a}
\ee

Upon quantization, we have
\be
 \left [ - \frac{m}{2} \Gam^\mu \Gam_\mu 
 - \frac{\Pi^\mu \Pi_\mu}{4 \mu}
  + \frac{1}{2m} (p^2-m^2)-\frac{Q}{2} \right ]  |\psi \rangle = 0 ,
\lbl{4.67}
\ee
where $\Gam^\mu$, $\Pi^\mu$ are the operators, satisfying
(\ref{4.11}),(\ref{4.12}), whereas $p_\mu$ is now the momentum operator that
commutes with $\Gam^\mu$, $\Pi^\mu$ and can be represented
as $p_\mu = - i \p/\p x^\mu$.
By using Eqs.\,(\ref{4.53}), the equation of state (\ref{4.67}) becomes
\be
  \left [-\frac{(D-1)m^2}{8 \mu (m^2-p^2)}+\frac{1}{2m}(p^2-m^2)-
  \frac{Q}{2} \right ] |\psi \rangle = 0.
\lbl{4.68}
\ee
Multiplication from the left by $2 m (m^2-p^2)$ gives
\be
  \left [-\frac{(D-1)m^3}{4 \mu}-(m^2-p^2)^2- m (m^2-p^2) Q \right ]
  |\psi \rangle = 0.
\lbl{4.69}
\ee
This is the 4th order equation in the derivatives with respect to spacetime
coordinates $x^\mu$, $\mu=0,1,2,3$. Additionally, we also have the relation
(\ref{4.54}), and consequently the equation of state (\ref{4.69}) becomes
\be
  \left [-\frac{(D-3)m^3}{4 \mu}- m \sqrt{\frac{m}{2 \mu}} Q \right ]
  |\psi \rangle = 0.
\lbl{4.70}
\ee
from which we obtain
\be
   Q= \frac{D-3}{2} \sqrt{\frac{m}{2 \mu}}\, = \,\frac{(D-3) \omega}{2} .
\lbl{4.71}
\ee
Then the term $Q/2$ that comes from the $(D+2)^{\rm th}$ time-like dimension
cancels the residual term (\ref{4.56}), occurring in Eq,\,(\ref{4.51}). 
The same cancelation happens if instead of introducing the extra variable $X^{D+2}$,
satisfying Eq.\,(\ref{4.61a}), we remain with $D+1$ variables, such that one
of them satisfies the analogous equation. Namely, if $D$ is any dimension,
then, of course, introducing an extra time-like dimension is equivalent
to assuming that three, and not only two of the $D+1$ dimensions are time-like,
so that, for instance, $X^{D-1} = K^2 \tau$, and ${\hat \mu}=0,1,2,...,D-2,D+1$.

Because Eq.\,(\ref{4.70}) is of such a simple form with only the $c$-numbers
within the bracket, the equation that determines $|\psi \rangle$ is in fact
(\ref{4.54}). This is the Klein-Gordon equation with the squared mass
$m^2 - m \sqrt{m/2 \mu}$. But because our physical states describe spin
$1/2$ particles, they must satisfy, not only the Klein-Gordon, but also
the Dirac equation
\be
  (i \gam^\mu \p_\mu - m_e ) |\psi \rangle~,~~~~
  m_e = \left (m^2 - m \sqrt{\frac{m}{2 \mu}} \right )^{1/2},
\lbl{4.72}
\ee
that is the``square root" of of Eq.\,(\ref{4.54}). The above equation (\ref{4.72}),
and consequently its ``square" (\ref{4.54}), is a condition that has to be
imposed in order to obtain the matching between the two different expressions,
(\ref{4.20}) and (\ref{4.21}) for the spin operator. With (\ref{4.72}), the
circle is closed, and the representation of operators $\Gam^\mu$, $\Pi^\mu$ in
terms of the Clifford numbers according to Eqs.\,(\ref{4.15}),(\ref{4.17}),
(\ref{4.25a}),(\ref{4.26}), is consistent with
the commutations relations (\ref{4.4})--(\ref{4.10a}), the definition
of spin operator (\ref{4.10b}), and also with the equation of state
(\ref{4.67}).

\section{Conclusion}

We have investigated an open string living in a target space with 
an extra time-like dimension. As an approximation, more precisely, as
a quenched description,  we obtained a point particle with
extrinsic curvature, which is responsible for the particle's spin.
The dynamics of such a system implies two first class and four second
class constraints. We quantized this system by employing the Dirac
brackets. We arrived at a system of operator equations that can be
satisfied by Clifford numbers expressed in terms of the Dirac gammas and
the generators of the Clifford algebra $Cl(0,2)$. The spin operator
of such quantized particle is expressed in terms of a superposition of the
commutators $[\gam^\mu,\gam^\nu]$, the coefficients being the projectors
onto the $(D-1)$-dimensional hypersurface, orthogonal to the particle's
momentum $p_\mu$, $\mu=0,1,2,...,D-1$.
It turns out that the Pauli-Lubanski operator, $S^\mu$, calculated from the
spin operator, is the same as that for the Dirac particle. The eigenvalues
of $S^\mu S_\mu$ are $s(s+1)=3/4$. This means that our quantized system has
spin $s=1/2$.

One has to take into account also the first class constraints.
Upon quantization, they become equations for physical states. 
But it turns out that with the Clifford algebra representation of
operators the condition $\phi_1 |\Psi \rangle = 0$ on physical states is not
satisfied, unless the dimension of spacetime is $D=3$. In order to
render the equation of state consistent for $D=4$ or higher, one has to bring
into the game one more extra time-like dimension, besides the two ones that are
already present in our model. Altogether, if we take $D=4$,
we have thus three time like and three space like dimensions, i.e., an
ultrahyperbolic space with neutral signature. Despite the fact that such spaces
for certain well known reasons are considered
as problematic for physics, there are works in the
literature\,\ci{UltraHyperbolicOK}--\ci{PavsicBook}
which reveal just the contrary.

A remarkable feature of the model presented in this paper  is that we started
from a bosonic string, and then described the motion of its end at $\sigma=0$ as
a point particle with extrinsic curvature (a variant of the so called `rigid
particle').
Upon quantization, we obtained a spin 1/2 system. This is a consequence of
another remarkable feature, namely that the algebra of the Dirac brackets
becomes upon quantization the algebra of commutation relations between the
operators that can be represented in terms of the gamma matrices.
The spin operator, defined as the commutator of Clifford numbers, matches
the spin operator, defined in terms of the velocity operator and its conjugate
momentum, if the states satisfy the Dirac equation. That extended objects, even
if apparently ``bosonic", can contain spinors, and thus spin one half
states, is a result implied more or less explicitly in many
works\,\ci{ExtendedClifford,ExtendedSpin}.
Those approaches are based on the fact that the extended objects can be sampled
by the center of mass coordinates, and the higher grade coordinates, describing
the area, 3-volume, 4-volume,.., associated with the object. Such a description
can be cast into an elegant form by means of Clifford
algebras\,\ci{ExtendedClifford}, and leads to the concept of Clifford space.
It is well known since Cartan\,\ci{Cartan}--\ci{CliffSpinors1}  that spinors are particular
Clifford numbers.
The fact that spinors are embedded in Clifford algebras has been explored in the
literature in various contexts\,\ci{CliffSpinors2}. In the present paper we
also sampled an extended object, but not in terms of Clifford numbers, but
in terms of the variables $x^\mu (\tau)$ that describe the motion of one
of the string ends on a surface $V_{D-1}$, and the variables
$y(\tau)$ that, up to the first order in the expansion (\ref{2.4}),
 describe the string's extension into the direction orthogonal
to $V_{D-1}$. Upon quantization, we
arrived at the spin one half states. This is in agreement with what comes out
from the Clifford algebra based approach to the extended objects. Because
our model of the type 2a rigid particle and its quantization involves
Clifford numbers, and because it is closely related to strings, 
it provides an additional test of the validity of the Clifford algebra
description of strings.
Moreover, we have confirmed and further elaborated  an
observation by some researchers\,\ci{Corben}--\ci{Kosyakov}
that the classical systems with spin resemble very closely the Dirac particle.
There is a number of other works that consider various approaches to
the classical particles with spin and their relation to the Dirac
equation\,\ci{SpinOthers}.

The connection of the type 2a rigid particle with strings, Clifford algebras
and Clifford spaces\,\ci{ExtendedClifford,ExtendedSpin,PavsicBook},
\ci{CliffUnific} which are very
promising in our attempts to construct
a unified theory of fundamental particles and interactions, including
quantum gravity, makes the results of this paper a further step
on our road towards the unification of the so far separate pieces of physics.
In this paper we considered the string with a finite length $L$, expanded
around $\sigma=0$. As an approximation we obtained the  action for
the point particle with curvature, where the terms of the order
$k^2 L^2$ and higher, that would give finer modulation of the trajectory
$x^\mu (\tau)$,  have been neglected. An interesting feature of
the latter action is that it possess non trivial limit $L \rightarrow 0$,
i.e., $m=LT \rightarrow 0$,
such that the term with curvature is still present. This means that if the
string length $L$ is decreasing, the string is becoming more and more
like a rigid particle. In the limit in which the string is infinitely short,
it behaves as the $m=0$ rigid particle, considered
by McKeon\,\ci{McKeon}.
We have touched this topics only in Footnote 2, whereas a detailed
investigation of the dynamics of such a particle has been considered
in Ref.\,\ci{ZeroLengthString}.

\vs{3mm}

\centerline{\bf Acknowledgement}

\vs{1mm}

This work has been supported by the Slovenian Research Agency.

\section*{Appendix: The Laplace equation for a time-like string and its
boundary conditions}

Let us investigate in more detail the situation of a string whose ends
are attached to two $Dp$-branes that sweep the wordlvolumes $V_D$ and $V'_D$,
as illustrated in Fig.\,1.

In the case of a usual, {\it space-like string} ($\epsilon = -1$), sweeping a
worldsheet with signature $(+-)$,  the string's embedding functions
$X^\mu (\tau,\sigma)$ satisfy in the conformal gauge the Helmholtz
equations of motion, subjected to a combination of Von Neumann and
Dirichlet boundary conditions, so that the string ends in general move
on a $D$-brane which in the case considered in Fig.\,1, is $Dp$-brane,
with $p=D-1$, sweeping the worldvolume $V_{D1}$.
The $(D+1)^{\rm th}$ dimension, along which the string extends, is space-like.

The situation is different in the case of a {\it time-like string} ($\epsilon =1$),
which corresponds to the worldsheet signature $(++)$ and time-like $X^{D+1}$,
then in the conformal gauge we have the following
equations of motion and the constraints:
\be
    {\ddot X}^{\hmu} + X''^\hmu = 0,~~~{\dot X}^\hmu {\dot X}_\hmu -
    X'^\hmu X'_\hmu = 0,~~~ {\dot X}^\hmu X'_\hmu = 0.
\lbl{A2}
\ee
Each embedding function thus satisfies the Laplace equation, subjected
to the above constraints. A general solution is
\bear
    &&X^\mu = C^\mu \tau + \sum_{n} (a_n^\mu {\rm cos}\, \omega_n \tau 
      + b_n^\mu {\rm sin} \, \omega_n \tau)
      (A_n {\rm e}^{k_n \sigma} + B_n {\rm e}^{-k_n \sigma})\nonumber \\
    &&X^{D+1} = \sigma, ~~~\sigma \in [0,L],
\lbl{A3}
\ear
where
\be
     \omega_n^2 - k_n^2 = 0, ~~~a_n^2 = b_n^2, ~~~C_\mu a_n^\mu 
     =C_\mu b_n^\mu = a_n^\mu {b_n}_\mu = 0, ~~~C^2=1.
\lbl{A4}
\ee
A particular solution is determined if the values of the functions $X^\hmu (\tau,\sigma)$
 on the boundary are given (Dirichlet boundary conditions):
\bear
  &&X^\mu (\tau,0) = x^\mu (\tau)~,~~~~~X^{D+1}(\tau,0) = 0,\lbl{A2a}\\
  &&X^\mu (\tau,L) = g^\mu (\tau)~,~~~~~X^{D+1}(\tau,L) = L,\lbl{A2b}\\
  &&X^\mu (\tau_1,\sigma) = F^\mu (\sigma),~~~~~~X^{D+1} (\tau_1,\sigma) = \sigma,
  \lbl{A2c}\\
  &&X^\mu (\tau_2,\sigma) = G^\mu (\sigma)~,~~~~~X^{D+1}(\tau_2,\sigma) = 
  \sigma. \lbl{A2d}
\ear

Now a question arises as to how an observer could manage such boundary
conditions on the branes $V_D$ and $V'_D$. Mathematically, of course we can
say that the string's worldsheet $V_2$ satisfies such boundary conditions and
the equations of motion (\ref{A2}).
But an observer, like us, cannot see the data in the entire space $M_{D+1}$.
According to our brane world scenario, the observers live in one of the
two branes, or both. Let us assume that the observers live in the brane $V_D$ at
$\sigma=0$, and that they  can only observe and determine the data in $V_D$.
 An observer in $V_D$ can thus only trace the functions
$X^\mu (\tau,0)=x^\mu (\tau)$, but not the entire string. From the point of
view of such an observer, functions $x^\mu (\tau)$ could be dynamical variables, if they
satisfied certain equations of motion. We have seen that, starting
from the action (\ref{2.1}), we can indeed derive dynamical equations
of motion for $x^\mu (\tau)$. This can be done if we expand the worldsheet
embedding functions $X^\mu (\tau,\sigma)$ into a Taylor series around
the point $\sigma =0$, according to Eq.\,(\ref{2.4}).

In Eqs.\,(\ref{A2})--(\ref{A2d}) we worked in a particular gauge, namely the conformal
gauge, whereas  in the derivation from Eq.\,(\ref{2.1}) to Eq.\,(\ref{2.15}), we did not
specify a gauge. A gauge is determined by a choice of the Lagrange multipliers
$e$ and $f$. If $e=1$ and $f=0$, then from Eqs. (\ref{2.12})--(\ref{2.14}) we have
$\gam^{11}= \frac{1}{\sqrt{\gam}}$, $\gam^{22}=\frac{1}{\sqrt{\gam}}$, $\gam^{12}=0$, which
means that the metric is conformal. In conformal gauge, the equations of motion
${\dot p}_\mu=0$ with $p_\mu$ given in Eq.\,(\ref{3.4}) become
\be
   \stackrel{ . . . .}{x}^\mu + \omega^2 {\ddot x}^\mu =0~, ~~~~~~
   \omega = \sqrt{\frac{m}{2 \mu}} = \frac{2}{L}.
\lbl{B1}
\ee
A general solution is
\be
  x^\mu (\tau) = C^\mu \tau + a^\mu {\rm cos} \, \omega \tau + b^\mu {\rm sin}\,\omega \tau .
\lbl{B2}
\ee
Using (\ref{2.20a}), we find
\be
  y^\mu (\tau) = \frac{L}{2 k} {\ddot x}^\mu = \frac{\omega^2 L}{2 k}
  (a^\mu {\rm cos} \, \omega \tau + b^\mu {\rm sin}\,\omega \tau ).
\lbl{B3}
\ee
Since $x(\tau) = X^\mu (\tau,0)$ and $y^\mu (\tau) = \frac{1}{k} X'^\mu (\tau,0)$, we
see that Eqs.\,(\ref{B2}) and (\ref{B3}) provide  boundary conditions at
$\sigma=0$ for the general solution (\ref{A3}). From Eq.\,(\ref{A3}) we have
\bear 
   &&X^\mu (\tau,0) = C^\mu \tau + \sum_{n} (a_n^\mu {\rm cos}\, \omega_n \tau 
      + b_n^\mu {\rm sin} \, \omega_n \tau)
      (A_n  + B_n ) \lbl{B4},\\
    &&X'^\mu (\tau,o) =  \sum_{n} (a_n^\mu {\rm cos}\, \omega_n \tau 
      + b_n^\mu {\rm sin} \, \omega_n \tau) k_n (A_n  - B_n ) .\lbl{B5}
\ear
This coincides with Eqs.\, (\ref{B2}) and (\ref{B3}), if $k_1=k=\omega = 2/L$, 
 $A_1 =1$, $a_1^\mu = a^\mu,~b_1^\mu=b^\mu$ and
$A_n =0$, $a_n^\mu = 0$, $b_n^\mu =0$ for $n \neq 1$.
The general solution in the presence of the boundary conditions
(\ref{B2}),(\ref{B3}) thus becomes
\bear
    &&X^\mu = C^\mu \tau + (a^\mu {\rm cos}\, \omega \tau 
      + b^\mu {\rm sin} \, \omega \tau) {\rm e}^{k \sigma}, \nonumber \\
    &&X^{D+1} = \sigma, ~~~\sigma \in [0,L] .
\lbl{B6}
\ear
It is illustrative to plot the solution (\ref{B6}). For this aim let us take
$C^\mu =(1,0,0,0) $, $a^\mu =(0,1,0,0)$, $b^\mu =(0,0,1,0)$, $\omega =1$.
We obtain a helix like solution given in Figs.\,2 and 3 for two different
values of $L$.

\setlength{\unitlength}{.8mm}
\begin{figure}[h!]
\hs{3mm}
\begin{picture}(100,80)

\put(25,-15){\includegraphics[scale=0.4]{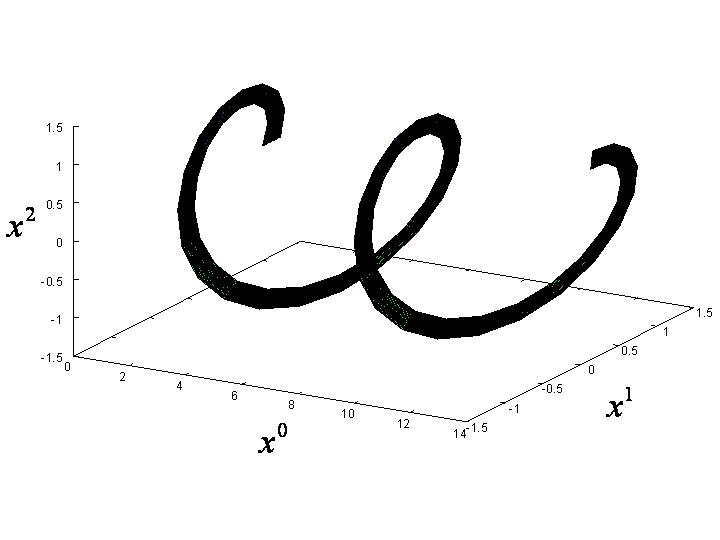}}
\end{picture}

\caption{\footnotesize The projection of the 2-dimensional worldsheet,
given by Eq.\,(\ref{B6}), onto the subspace, described by coordinates
$(x^0,x^1,x^2)$, if the string is short. We see that the boundary curve
for such a short string is a good approximation to the string worldsheet.}

\end{figure} 

\setlength{\unitlength}{.8mm}
\begin{figure}[h!]
\hs{3mm}
\begin{picture}(100,75)

\put(25,-17){\includegraphics[scale=0.4]{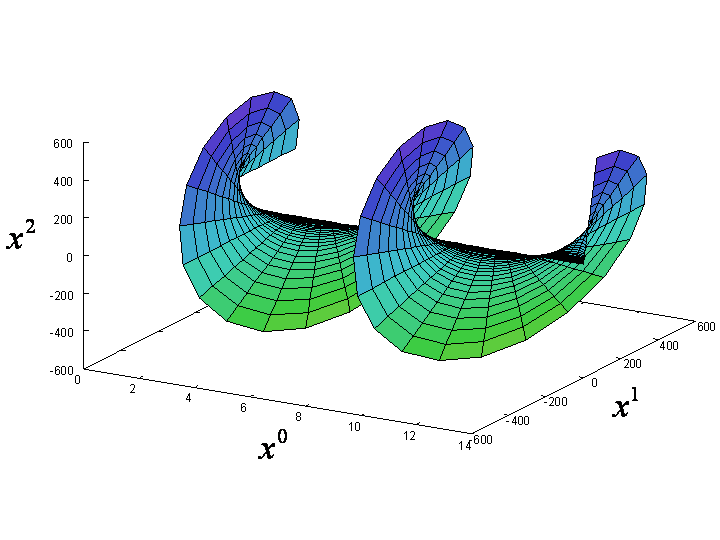}}

\end{picture}

\caption{\footnotesize The projection of the 2-dimensional worldsheet,
given by Eq.\,(\ref{B6}), onto the $(x^0,x^1,x^2)$-subspace, if
the string is long.}

\end{figure} 

By solving the equations of motion (\ref{B1}),(\ref{B3}) for $x^\mu (\tau)$ and
$y^\mu (\tau)$, we have thus found boundary conditions for the
string that satisfies the equations of motion (\ref{A2}).  A boundary at
$\sigma=0$ of our time-like string is the point particle with
extrinsic curvature. Of course, this is not the only possible boundary.
If we took more terms in the expansion (\ref{2.4}), then we would obtain the
equations of motion for more terms $y_i^\mu (\tau)$, whose solution
would be a more complicated curve than the boundary $\sigma=0$
in Eq.\, (\ref{B6}), and a more complicated wordlsheet than in Fig.\,2 or Fig.\,3.

Finally, let me comment on the boundary conditions that an open
string, either space-like or time-like has to satisfy. First it was
believed that an open (space-like) string must satisfy the Von
Neumann boundary conditions. Later it was found  that it
can satisfy the Dirichlet boundary conditions, as well  as
a combinations of those two types of conditions, which led
to the discovery of $D$-branes.

This suggests a thesis that any possible solution of
the string equations of motion can in principle be realized
in Nature. One should not impose certain boundary conditions
and consider only those solutions that satisfy them. Every
solution corresponds to a possible physical situation related
to how the string ends are coupled (or not coupled) to other
physical objects (e.g., strings and branes). In a quantized
theory, all those possibilities have to be taken into account
in the wave functional. In fact, this is the essence of quantum
theory. While in the classical theory of a point particle there
are differential equations of motion that have a class of
possible solutions, and one has to choose one particular
solution by imposing some initial conditions, in the
quantum theory all those possibilities are contained in the wave
function. Analogous must hold for extended objects, such
as strings that satisfy partial differential equations. They
allow for a class of possible solutions, and one has to
specify initial and boundary conditions in order to obtain
one particular solution.  Usually, in the quantized string theory
all possible solutions corresponding to all possible
initial conditions are taken into account, but there is
still a restriction due to choice of boundary conditions.
In a complete quantization, one should take into account
all possible boundary conditions as well, and thus the total
class of possible solutions. This does not mean that a 
``partial" quantization of the string theory is not correct,
it only does not take into account the whole story.
An important step forward into this direction was the
the discovery of $D$-branes as dynamical objetcs.

In the case of a time-like string that satisfies the Laplace equation
(\ref{A2}), the situation is  more straightforward, because here the Laplace
equation is not, like in electrostatics, a static case of a
Helmoltz equation, and therefore there is no need to care
about the momentum transfer across the string ends. All directions
along the string worldsheet  are time-like. 
However, a more than one-dimensional time is not consistent
with the dynamics as we experience it. In order to to obtain a one
dimensional time, one has to intersect the two-time worldsheet $V_2$ with
a space-like surface $\Sigma\subset V_{D+1}$ and consider the dynamics
on $\Sigma$. It must be a consistent dynamics, with
conserved energy-momentum. In our scenario, the space-like
surface $\Sigma$ is the brane $V_{D-1}$  that does not intersect,
but touches $V_2$ at the string end and sweeps the worldvolume
 $V_D$. The dynamical equations of  the motion of the string end
at $\sigma=0$, considered in this paper, are the
equations for a point particle with extrinsic curvature, with
the conserved momentum $p_\mu$. They were obtained by
expanding the string embedding functions $X^\mu (\tau,\sigma)$ in terms
of the parameter $\sigma$ up to the first order. Inclusion of higher orders in the expansion
(\ref{2.4}) would modify the point particle action so that besides the extrinsic
curvature it would contain torsion and higher curvatures as well. In the
corresponding equations of motion not  only fourth, 
but also higher derivatives would enter the game. A solution of such
equations of motion would also be consistent with the boundary of
a time-like Nambu-Goto string. Regardless of how many terms we take in the
expansion of $X^\mu (\tau,\sigma)$, we always obtain an exact boundary
of the string, satisfying certain higher derivative equations of motion.
It is reasonable to anticipate that for a given dimension $D$ the expansion (\ref{2.4})
above certain order brings nothing new to the equations of motion
of the boundary point.  Namely, the expansion to the first order gives the point-particle
with extrinsic curvature, whereas the expansion to higher orders presumably gives
the point particle action with higher curvatures. Because according to the
Frenet formula in $D$-dimension there are $D-1$ generalized curvatures associated
with a curve, this imposes the limit to how many curvature terms can be in
the action for a point particle whose worldline traces the motion of the string end.
The reasoning along such lines thus leads to the point particle with $D-1$
generalized curvatures, where the point particle with extrinsic curvature, considered
in this paper, is just a special case.

\end{document}